\documentclass[iop,dvipdfmx]{emulateapj}
\usepackage{natbib,aas_macros,amsmath}

\newcommand{\average}[1]{\ensuremath{\langle#1\rangle} }

\begin{document}
\slugcomment{Accepted for publication in ApJ}
\shorttitle{Full-Data Results of Hubble Frontier Fields}
\shortauthors{Ishigaki et al.}

\title{
Full-Data Results of Hubble Frontier Fields: \\
UV Luminosity Functions at $z\sim6-10$ and a Consistent Picture of Cosmic Reionization
}

\author{
Masafumi~Ishigaki\altaffilmark{1,2}, 
Ryota~Kawamata\altaffilmark{3},
Masami~Ouchi\altaffilmark{1,4},
Masamune~Oguri\altaffilmark{2,4,5}, 
Kazuhiro~Shimasaku\altaffilmark{3,5},\\
and Yoshiaki~Ono\altaffilmark{1}
}
\email{ishigaki@icrr.u-tokyo.ac.jp}
\altaffiltext{1}{Institute for Cosmic Ray Research, The University of Tokyo, 5-1-5 Kashiwanoha, Kashiwa, Chiba 277-8582, Japan}
\altaffiltext{2}{Department of Physics, Graduate School of Science, The University of Tokyo, 7-3-1 Hongo, Bunkyo-ku, Tokyo 113-0033, Japan}
\altaffiltext{3}{Department of Astronomy, Graduate School of Science, The University of Tokyo, 7-3-1 Hongo, Bunkyo-ku, Tokyo 113-0033, Japan}
\altaffiltext{4}{Kavli Institute for the Physics and Mathematics of the Universe (Kavli IPMU, WPI), The University of Tokyo, 5-1-5 Kashiwanoha, Kashiwa, Chiba 277-8583, Japan}
\altaffiltext{5}{Research Center for the Early Universe, The University of Tokyo, 7-3-1 Hongo, Bunkyo-ku, Tokyo 113-0033, Japan}

\begin{abstract}
We present UV luminosity functions of dropout galaxies at $z\sim6-10$ with
the complete Hubble Frontier Fields data.
We obtain a catalog of $\sim450$ dropout-galaxy candidates 
(350, 66, and 40 at $z\sim6-7$, 8, and 9, respectively), whose
UV absolute magnitudes reach $\sim-14$ mag, $\sim2$ mag deeper than the Hubble Ultra Deep Field
detection limits. We carefully evaluate number densities of the dropout galaxies by Monte-Carlo simulations,
including all lensing effects such as magnification, distortion, and multiplication of images
as well as detection completeness and contamination effects in a self-consistent manner.
We find that UV luminosity functions at $z\sim6-8$ have steep faint-end slopes, $\alpha\sim-2$, 
and likely steeper slopes, $\alpha\lesssim-2$ at $z\sim9-10$.
We also find that the evolution of UV luminosity densities shows a non-accelerated decline 
beyond $z\sim8$
in the case of $M_\mathrm{trunc}=-15$, while an accelerated 
in the case of $M_\mathrm{trunc}=-17$.
We examine whether our results are consistent with
the Thomson scattering optical depth from the Planck satellite
and the ionized hydrogen fraction $Q_\mathrm{HII}$ at $z\lesssim7$
based on the standard analytic reionization model.
We find that there exist reionizaiton scenarios that consistently explain all the observational
measurements with the allowed parameters of $f_{\rm esc}=0.17^{+0.07}_{-0.03}$ and $M_\mathrm{trunc}>-14.0$
for $\log\xi_\mathrm{ion}/[\mathrm{erg}^{-1}\ \mathrm{Hz}]=25.34$, where
$f_{\rm esc}$ is the escape fraction, 
$M_\mathrm{trunc}$ is the faint limit of the UV luminosity function,
and $\xi_\mathrm{ion}$ is the conversion 
factor of the UV luminosity to the ionizing photon emission rate.
The length of the reionization period  
is estimated to be $\Delta z=3.9^{+2.0}_{-1.6}$ (for $0.1<Q_\mathrm{HII}<0.99$),
consistent with the recent estimate from Planck.
\end{abstract}

\keywords{
galaxies: formation --
galaxies: evolution --
galaxies: high-redshift
}

\section{Introduction} \label{sec:Introduction}

Understanding the sources of cosmic reionization is one of the goals of modern astronomy.
Observational studies show that the reionization occurred at $z\gtrsim6$ \citep{fan2006a,pentericci2011,pentericci2014,ono2012,totani2006,totani2014,kashikawa2011,ouchi2010,konno2014,treu2013}.
Major sources of the cosmic reionization are thought to be star-forming galaxies. 
Recent studies indicate that UV luminosity functions of the star-forming galaxies have steep faint-end slopes at $z\gtrsim6$ \citep{bouwens2015,schenker2013,mclure2013}, which suggests that the ionizing photons are mainly supplied by faint galaxies.
However, high-$z$ galaxy observations, such as Hubble Ultra Deep Field, can reach the limiting absolute magnitudes down to $\lesssim -16$ mag at $z\gtrsim6$.
The abundance of galaxies is not well known at the magnitude fainter than $\sim-16$ where contributions to ionizing photon budget may be dominated.
Moreover, there exist large uncertainties of various parameters to estimate an ionizing photon emission rate from the galaxy abundance obtained by observations. 
One of the important parameters is the numerical factor $\xi_\mathrm{ion}$ that converts a UV luminosity density to the ionizing photon emission rate of a star-forming galaxy.
\citet{bouwens2016c} estimated that $\log \xi_\mathrm{ion}/[\mathrm{erg}^{-1}\ \mathrm{Hz}]$ has a mean value of $25.34\pm0.02$, albeit at $z\sim4-5$, the post-reionization epoch.
(see also \citet{nakajima2016} for the recent studies of $\xi_\mathrm{ion}$).
Another important parameter is the escape fraction of ionizing photons $f_\mathrm{esc}$, which is the fraction of the number of escaping ionizing photons to that of ionizing photons produced in a galaxy.
\citet{steidel2001} estimated that the average value of $f_\mathrm{esc}$ is $\sim0.1$ at $z\sim3$ (see also \citealt{shapley2006,iwata2009,nestor2011,japelj2016}).
\citet{ono2010} investigated the stellar populations of Ly$\alpha$ emitters at $z\sim6$, and obtained weak upper limits of $f_\mathrm{esc} \sim 0.6$ at $z=5.7$ and $f_\mathrm{esc}\sim0.9$ at $z=6.6$, based on the constraints on the nebular emission line fluxes. 
Although many constraints are given by these studies, reasonable constraints on $\xi_\mathrm{ion}$ and $f_\mathrm{esc}$ at the epoch of reionization have not been obtained so far.

Gravitational lensing is an effective tool for studying high-redshift galaxies.
Cluster Lensing And Supernova survey with Hubble (CLASH; \citealt{postman2012}) has discovered many candidates of faint galaxies at $z\gtrsim6$ strongly lensed by galaxy clusters.
In October 2013, Hubble Frontier Fields (HFF; PI: Lotz) program has started observing six massive clusters: Abell 2744, MACSJ0416, MACSJ0717, MACSJ1149, AbellS1063, and Abell370 \citep{lotz2016}.
The HFF program achieves the limiting magnitudes $\sim29$ AB mag at the $5\sigma$ level, which are $\sim1$ mag deeper than those of CLASH.
Many groups study the properties of high-redshift galaxies using a part of the HFF data \citep{atek2015a,atek2015b,zheng2014,coe2015,oesch2015,ishigaki2015,ishigaki2016,kawamata2015,kawamata2016,livermore2016,bouwens2016,bouwens2016b,mcleod2015,mcleod2016}.
In September 2016, the observations for all of the six HFF clusters are completed.
Here we exploit the full six cluster HFF data to investigate the galaxies at the reionization epoch.

In this paper, we first present details of the HFF data in Section \ref{sec:Data}.
In Section \ref{sec:Samples}, we select high-redshift galaxies with the dropout selection technique.
We obtain UV luminosity functions in Section \ref{sec:UV_LF}, and discuss the properties of the faint galaxies in Section \ref{sec:Discussion}.
Finally, we summarize our results in Section \ref{sec:Summary}.
Throughout this paper, we use a cosmology with $\Omega_\mathrm{m} = 0.3$, $\Omega_\mathrm{\Lambda} = 0.7$, $\Omega_\mathrm{b} = 0.04$, and $H_0 = 70$ km s$^{-1}$ Mpc$^{-1}$.

\section{Data} \label{sec:Data}

For determination of UV luminosity functions, 
we use the complete samples of the HFF fields; Abell 2744, MACSJ0416, MACSJ0717, MACSJ1149,  AbellS1063, and Abell370.
Cluster and parallel fields of the six massive clusters have been observed with Advanced Camera for Surveys (ACS) and Wide Field Camera 3 (WFC3) of Hubble Space Telescope (HST) through 2013 October to 2016 September.
The total survey area covers $\sim56$ arcmin$^2$.
Drizzled and weight images are produced and released by the Space Telescope Science Institute (STScI) at the HFF Web site.\footnote{http://www.stsci.edu/hst/campaigns/frontier-fields/}
The images consist of three ACS bands and four WFC3-IR bands; F435W ($B_{435}$), F606W ($V_{606}$), F814W ($i_{814}$), F105W ($Y_{105}$), F125W ($J_{125}$), F140W ($JH_{140}$), and F160W ($H_{160}$).

First, we homogenize the Point Spread Function (PSF) of the WFC3 images.
The PSF FWHM of homoginized images is $\sim 0.18$ arcsec.
We then measure the limiting magnitudes. 
We divide each image into $4\times4$ grid cells and define the limiting magnitude in each cell. 
This is because limiting magnitudes are not homogeneous due to the intra cluster light. 
The $5\sigma$ limiting magnitudes in the $H_{160}$ band images are $\sim 28.5 - 29$ mag in a $0\farcs35$-diameter circular aperture.
Details of the PSF homogenization procedure and the limiting magnitude measurements are described in \citet{kawamata2016} (see also \citealt{ishigaki2015}).
A summary of the observational properties is provided in Tables \ref{table:summary} and \ref{table:limitmag}.

\begin{deluxetable*}{lcccccc}
\tablecaption{Observational properties and the numbers of high-$z$ galaxy candidates of the HFF Data\label{table:summary}}
\tablewidth{0pt}
\tablehead{
		\colhead{Field} & \colhead{Depth ($H_{160}$)\tablenotemark{a}} & \colhead{Area\tablenotemark{b}}  & \colhead{Number of Galaxies} & \colhead{} & \colhead{} & \colhead{}\\
		\colhead{} & \colhead{($5\sigma$ limit mag)} & \colhead{(arcmin$^2$)} & \colhead{$z\sim6-7$} & \colhead{$z\sim8$} & \colhead{$z\sim9$} & \colhead{$z\sim10$}   
}
\startdata
Abell2744 Cluster	&	28.88	&	4.75 (1.52\tablenotemark{c})	&	22	&	10	&	3	&	0	\\
Abell2744 Parallel	&	29.11	&	4.32	&	30	&	4	&	4	&	0	\\
MACS0416 Cluster	&	28.84	&	4.73 (2.00\tablenotemark{c})	&	25	&	5	&	3	&	0	\\
MACS0416 Parallel	&	29.30	&	4.78	&	23	&	9	&	6	&	0	\\
MACS0717 Cluster 	&	28.56	&	4.58 (0.75\tablenotemark{c})	&	22	&	0	&	0	&	0	\\
MACS0717 Parallel	&	28.88	&	4.25	&	41	&	6	&	3	&	0	\\
MACS1149 Cluster 	&	28.87	&	4.66 (1.32\tablenotemark{c})	&	41	&	2	&	4	&	0	\\
MACS1149 Parallel	&	29.04	&	4.72	&	37	&	10	&	3	&	0	\\
AbellS1063 Cluster  &	28.85	&	4.25 (0.98\tablenotemark{c})	&	23	&	4	&	2	&	0	\\
AbellS1063 Parallel &	29.12	&	4.36	&	39	&	6	&	3	&	0	\\
Abell370 Cluster    &	28.65	&	4.14 (0.73\tablenotemark{c})	&	7	&	6	&	2	&	0	\\
Abell370 Parallel   &	29.14	&	4.31	&	40	&	3	&	7	&	0	\\
\tableline 
Total   &		&		&	350	&	66	&	40	&	0

\enddata
\tablenotetext{a}{The depths are defined with a $0\farcs35$-diameter aperture.}
\tablenotetext{b}{The area of each image is reduced by the masking near the bright stars.}
\tablenotetext{c}{The survey area in the source plane at $z=7$.}
\end{deluxetable*}

\begin{deluxetable*}{lccccccc}
\tablecaption{$5\sigma$ limiting magnitudes\label{table:limitmag}}
\tablewidth{0pt}
\tablehead{
		\colhead{Field} & \colhead{$B_{435}$} & \colhead{$V_{606}$} & \colhead{$i_{814}$} & \colhead{$Y_{105}$} & \colhead{$J_{125}$} & \colhead{$JH_{140}$} & \colhead{$H_{160}$} 
}
\startdata
Abell2744 Cluster	&	28.72	&	28.83	&	28.94	&	29.20	&	28.87	&	28.92	&	28.88	\\
Abell2744 Parallel	&	28.84	&	29.20	&	29.03	&	29.39	&	29.00	&	29.03	&	29.11	\\
MACS0416 Cluster	&	28.74	&	29.08	&	28.96	&	29.11	&	28.85	&	28.87	&	28.84	\\
MACS0416 Parallel	&	28.65	&	28.92	&	29.01	&	29.45	&	29.22	&	29.26	&	29.30	\\
MACS0717 Cluster 	&	28.70	&	28.74	&	28.76	&	28.91	&	28.61	&	28.70	&	28.56	\\
MACS0717 Parallel	&	28.96	&	29.00	&	29.01	&	29.16	&	28.94	&	28.95	&	28.88	\\
MACS1149 Cluster 	&	28.57	&	28.89	&	28.94	&	29.17	&	28.81	&	28.73	&	28.87	\\
MACS1149 Parallel	&	28.50	&	28.83	&	28.90	&	29.35	&	28.98	&	29.06	&	29.04	\\
AbellS1063 Cluster  &	28.69	&	28.86	&	28.90	&	29.19	&	28.87	&	28.98	&	28.85	\\
AbellS1063 Parallel &	28.87	&	29.66	&	29.09	&	29.43	&	29.06	&	29.19	&	29.12	\\
Abell370 Cluster    &	28.64	&	28.78	&	28.86	&	29.00	&	28.69	&	28.74	&	28.65	\\
Abell370 Parallel   &	28.65	&	28.90	&	29.07	&	29.45	&	29.00	&	29.01	&	29.14	
\enddata
\end{deluxetable*}

\section{Samples} \label{sec:Samples}
In this study, we select $i$-dropout ($z\sim6-7$), $Y$-dropout ($z\sim8$), $YJ$-dropout ($z\sim9$), and $J$-dropout ($z\sim10$) galaxy candidates.
A detailed description of the dropout selections is given by \citet{kawamata2016}.
In this section, we give a brief description about dropout selections.
First, we create detection images using {\sc SWarp} \citep{bertin2002}.
The dection images are combination of $J_{125}$, $JH_{140}$, and $H_{160}$ bands for the $i$- and $Y$-dropout selections, $JH_{140}$ and $H_{160}$ bands for the $YJ$-dropout selection, and $H_{160}$ band for the $J$-dropout selection, respectively.
We run SExtractor (version 2.8.6; \citealt{bertin1996}) in dual-image mode using the detection images, and create photometric catalogs.
The colors of galaxies are measured with {\tt MAG\_APER} $m_\mathrm{AP}$.
The total magnitudes of galaxies $m_\mathrm{tot}$ are defined with the equation $m_\mathrm{tot} = m_\mathrm{AP} - 0.82$, where the offset $0.82$ corresponds to the aperture correction and was derived in \citet{ishigaki2015}.
Selection criteria of $i$-dropout galaxies are
\begin{eqnarray}
i_{814} - Y_{105} > 0.8, \\
Y_{105} - J_{125} < 0.8, \\
i_{814} - Y_{105} > 2(Y_{105} - J_{125}) + 0.6,
\end{eqnarray}
which are used in \citet{atek2015b}.
We require the detection significance levels beyond $5\sigma$ level in the $Y_{105}$ band and $J_{125}$ band.
We exclude the objects that are detected at the $2\sigma$ level in both the $B_{435}$ and $V_{606}$ bands or in the $B_{435}+V_{606}$ stacked images.

For $Y$-dropouts, we adopt the selection criteria given by \citet{atek2014}:
\begin{eqnarray}
Y_{105} - J_{125} > 0.5, \\
J_{125} - JH_{140} < 0.5, \\
Y_{105} - J_{125} > 1.6(J_{125} - JH_{140}) + 0.4.
\end{eqnarray}
Similarly, we apply the source detection thresholds of the $>5\sigma$ significance levels in the $J_{125}$, $JH_{140}$, and $H_{160}$ bands.
We remove the objects that are detected at the $2\sigma$ level in at least one of the $B_{435}$, $V_{606}$, or $i_{814}$ bands.

For $YJ$-dropouts, we use criteria presented in \citet{oesch2013} and \citet{ishigaki2015}:
\begin{eqnarray}
(Y_{105} + J_{125})/2 - JH_{140} > 0.75, \\
\begin{split}
(Y_{105} + J_{125})/2 - JH_{140} > \\
0.8(JH_{140} - H_{160}) + 0.75, 
\end{split} \\
J_{125} - H_{160} < 1.15, \\
JH_{140} - H_{160} < 0.6.
\end{eqnarray}
We require the detection significance levels beyond $3\sigma$ level in both the $JH_{140}$ and $H_{160}$ bands
and $3.5\sigma$ level in one of the $JH_{140}$ or $H_{160}$ bands.
Again, we remove the objects that are detected at the $2\sigma$ level in the $B_{435}$, $V_{606}$, or $i_{814}$ bands.

For $J$-dropouts, we apply the criteria used in \citet{oesch2015}:
\begin{eqnarray}
J_{125} - H_{160} > 1.2, \\
S/N(B_{435}\mathrm{\ to\ }Y_{105}) < 2.
\end{eqnarray}
We also require $\chi^2_{opt+Y} < 2.5$, 
where $\chi^2_{opt+Y}$ is defined by $\chi^2_{opt+Y} \equiv \sum_n \mathrm{SGN}(f_n)S/N(n)^2$.
$f_n$ is the $n$th band flux in $B_{435}$, $V_{606}$, $i_{814}$, and $Y_{105}$. $\mathrm{SGN}(f_n)$ is a sign function.
$S/N$s in the $JH_{140}$ and $H_{160}$ bands are required to be $>3.5\sigma$ in both bands
and $>5\sigma$ in one of the two bands.
Finally, spurious sources are removed from the dropout galaxy candidates by visual inspection.

In this study, we find no $J$-dropout candidates in the all cluster and parallel fields.
Our dropout samples consist of 350 $i$-dropouts, 66 $Y$-dropouts, and 40 $YJ$-dropouts in total.
The number of the dropout candidates in each field is listed in Table \ref{table:summary}.
We list these dropout candidates in Tables \ref{candidates7}, \ref{candidates8}, and \ref{candidates9}.
The values of magnitudes are slightly different from those in \citet{kawamata2016} because \citet{kawamata2016} used an older version of SExtractor.

\section{UV Luminosity Functions}\label{sec:UV_LF}

In this section, we calculate UV luminosity functions at $z\sim6-7$, $8$, $9$, and $10$.
UV luminosity functions are represented by the Schechter function
with three parameters; $\phi_\ast$, $M_\ast$, and $\alpha$:
\begin{eqnarray}
\begin{split}
\Phi(M_\mathrm{UV},\phi_\ast,M_\ast,\alpha)dM_\mathrm{UV} = 0.4\phi_\ast \ln(10) \\
\times \exp\left[10^{-0.4(M_\mathrm{UV}-M_\ast)}\right] 10^{-0.4(\alpha+1)(M_\mathrm{UV}-M_\ast)}dM_\mathrm{UV}.
\end{split}
\label{eq:schechter}
\end{eqnarray}
We derive the luminosity functions basically in the same manner as that described in Section 5 of \citet{ishigaki2015},
that adopt the luminosity function fitting on the image plane.
We refer to the method as the image plane method in the remainder of this paper.
The image plane method deals with not only the lensing magnification effects, 
but also the distortion and multiplication of lensed images in a self-consistent way.
\citet{oesch2015} showed that the detection completeness of galaxies strongly depends on the galaxy size and the image distortion,
especially at areas with high magnifications.
It is critical to properly evaluate the lensing effects and the properties of high-redshift galaxies 
to derive the luminosity functions, as done in this paper.
In this study, we define $M_\mathrm{UV}$ as the magnitude in the rest-frame wavelength of $1500$\AA.
This wavelength roughly corresponds to $J_{125}$, $JH_{140}$, $H_{160}$, and $H_{160}$ bands at $z\sim6-7$, $8$, $9$, and $10$, respectively.

In section \ref{sec:mock_catalogs}, we create mock catalogs of high-redshift galaxies and obtain number densities on the image plane.
In section \ref{sec:fitting}, we compare number densities of the mock catalogs with those of the observation, and determine the best fit Schechter parameters.

\subsection{Creation of Mock Catalogs}\label{sec:mock_catalogs}

We create mock catalogs of simulated high-redshift galaxies at $z\sim5.0-12$.
The number densities of the galaxies follow the Schechter functions.
The half-light radius of the galaxies are $\sim0.45$, $\sim0.22$, and $\sim0.11$ kpc for galaxies with magnitudes of $M_\mathrm{UV} < -20$, $-20 < M_\mathrm{UV} < -16$, and $M_\mathrm{UV} > -16$, respectively.
This assumption is consistent with the size--luminosity relations adopted in previous studies.
The S\'ersic index is assumed to be $1$.
The ellipticities are randomly chosen from the range of $0.0$--$0.9$,
because high-redshift galaxy studies show roughly uniform distributions of the ellipticities (e.g. \citealt{ravindranath2006}).
The UV spectral slope $\beta$ is assumed to be $\beta=-2$,
which is consistent with the result of \citet{bouwens2014}.
The UV spectra are attenuated with the IGM absorption model of \citet{madau1996}.
The mock galaxies are distributed in random positions of the source plane.
We create images of the mock galaxies with {\tt writeimage} function in the lens model software {\sc glafic} \citep{oguri2010}.
The {\tt writeimage} function calculates the gravitational lensing effects including magnification, distortion, and multiplication, for an object and produce the image(s) in the image plane.
Here we use the mass models of the six clusters described in \citet{kawamata2016} and \citet{kawamata2017}.
Three out of the six models are released in the HFF Web site\footnote{https://archive.stsci.edu/prepds/frontier/lensmodels/}
as version 3 for Abell 2744, MACS0717, and MACS1149.
We use an updated version of the mass model for MACS0416 \citep{kawamata2017}.
In the parallel fields, we use {\tt writeimage\_ori} function, which creates unlensed images.
We add the simulated images in the real HFF data.
Then we detect the mock galaxies with SExtractor and select dropouts in the mock catalogs with the color criteria same as the observations described in Section \ref{sec:Samples}.
We obtain number densities of dropouts in the mock catalogs.

Low-redshift galaxies are potentially contaminate the dropout catalogs of the observational data.
We estimate the contamination rate by the following procedure.
First, we create the catalogs of bright ($22 < H_{160} < 25$) objects in the HFF images.
In this magnitude range, all objects do not meet the color criteria of the dropouts, which are probably low-redshift interlopers.
Then we create interloper catalogs of faint objects ($25 < H_{160} < 29$).
We fit the number densities of the bright interlopers with a power-law function, 
and match the number densities of the faint interlopers to those extrapolated from the bright magnitudes.
We assume that the physical properties of faint interlopers such as the color and size are the same as those of the bright objects.
We create images of artificial objects from these catalogs with the {\tt mkobjects} package of {\sc iraf},
and randomly place them on the HFF images.
We detect the artificial objects with SExtractor,
and select the artificial objects that meet the dropout selection criteria.
We regard these selected objects as contaminants.
We thus obtain the contamination rate of dropouts as a function of magnitude.
The contamination rate is $\sim 9\%$ ($\sim 32\%$) in the magnitude range of $M_\mathrm{UV} < 27.25$ ($M_\mathrm{UV} > 27.25$) on average.
We add the number densities of the contaminants to the total number densities of the dropouts in the mock catalogs.

\subsection{Luminosity Function Fitting}\label{sec:fitting}

We compare the observed number densities of dropouts with the number density obtained in the simulations, which is described in the Section \ref{sec:mock_catalogs}.
We obtain the best-fit luminosity function by fitting these number densities with a maximum likelihood method.
We calculate the likelihood with the following equation, assuming the Poisson errors:
\begin{equation}
{\cal L} 
	\propto \prod_{\rm field} \prod_i 
		n_{{\rm sim},i}^{n_{{\rm obs},i}} 
		e^{-n_{{\rm sim},i}},
\end{equation}
where $n_{{\rm sim},i}$ is the simulated number counts in an $i$th magnitude bin,
and $n_{{\rm obs},i}$ is the observed number counts in the HFF in the magnitude bin.
We treat the three Schechter parameters, $\phi_\ast$, $M_\ast$, and $\alpha$, as free parameters in the fitting of luminosity functions at $z\sim6-7$ and $8$.
The number counts at $z\sim9$ and $z\sim10$ have too poor statistcis to constrain the three Schechter parameters.
Therefore, we also provide results where we treat $\phi_\ast$ as the only free parameter at $z\sim9$ and $10$, and
fix $M_\ast$ and $\alpha$ to the values at $z\sim8$, $-20.35$ and $-1.96$, respectively.
In order to improve the statistics, we simultaneously fit our number counts data and UV luminosity function data points from previous studies.
We use recent results of blank field surveys; \citet{bouwens2015,bowler2014,ouchi2009,bradley2012,oesch2013,calvi2016}.
Although \citet{Schmidt2014} presented an updated result of the BoRG survey \citep{bradley2012}, we do not incorporate their result because they do not provide number densities at individual magnitudes.
We iterate the luminosity function fitting with various Schechter parameter sets,
and obtain the best-fit parameters. 

One of the advantages of the image plane method is that the result is not affected by uncertainties of magnification factors of dropout candidates.
If a galaxy has magnification factor larger than $\sim100$, the $1\sigma$ error of the magnification is as large as itself (Tables 10 to 12 in \citealt{kawamata2016}).
Luminosity function fitting on the source plane has large uncertainty in the faint end,
because the galaxy samples in the faint magnitude bins consists of galaxies with high magnification factors.
The disadvantage of the image plane method is that there is a degeneracy between number densities of intrinsically bright and faint galaxies with the same apparent magnitude.
The constraints on the faint end of luminosity functions are weaker than those with fitting on the source plane.
In order to strengthen the constraints on the faint end,
we divide the dropout samples into subsamples with magnification factor binning of $\mu < 2$, $2 < \mu < 6$, $6 < \mu < 18$, and $\mu > 18$ at $z\sim6-7$. 
We use the surface number density in each magnitude bin and in each magnification factor bin for fitting.
Note that the number densities divided with magnification factor bins depend on mass models,
although the uncertainty is not larger than the one in fitting on the source plane.
We estimate the uncertainty of mass models in the end of this section.
At $z \gtrsim 8$, we do not divide the samples,
because the number of $z \gtrsim 8$ galaxy candidates is not large enough for making subsamples.

\begin{figure}
	\epsscale{1.2}
	\plotone{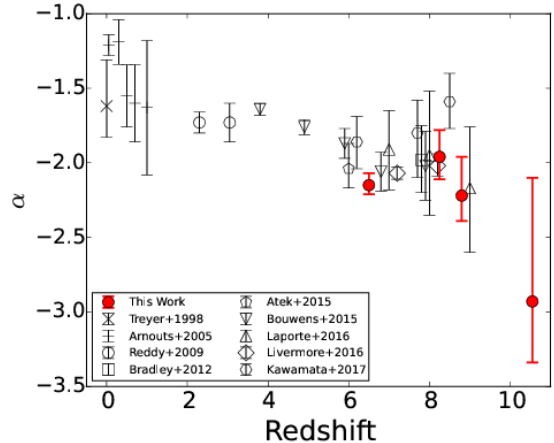}
	\caption{Redshift evolution of the faint-end slope $\alpha$ of the rest-frame UV luminosity function. The red circles show our results and the black symbols show the values of local UV-selected galaxies \citep{treyer1998}, $0.2\leq z \leq 1.2$ galaxies \citep{arnouts2005}, $1.9\leq z \leq 3.4$ LBGs \citep{reddy2009}, $z\sim8$ LBGs \citep{bradley2012}, $z\sim6$ LBGs \citep{atek2015a}, $z\sim4-8$ LBGs \citep{bouwens2015}, $z\sim7-9$ LBGs \citep{laporte2016}, $z\sim7-8$ LBGs \citep{livermore2016}, and $z\sim6-9$ LBGs \citep{kawamata2017}.
	}
	\label{fig:alpha}
\end{figure}

\begin{figure}
	\epsscale{1.2}
	\plotone{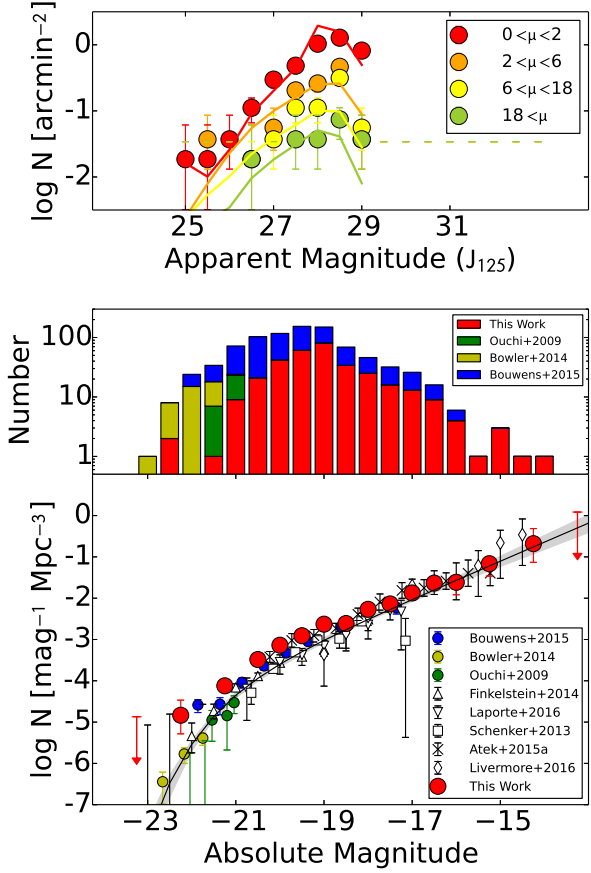}
	\caption{Surface number densities, histograms of number counts, and luminosity functions of $z \sim 6-7$ dropout candidates.
Top panel: surface number densities of our dropout candidates in the HFF data (circles) and those derived from the best-fit Schechter function (lines).
The colors of the circles and lines denote the magnification factor of $0<\mu<2$ (red), $2<\mu<6$ (orange), $6<\mu<18$ (yellow), and $18<\mu$ (green).
The horizontal axis shows the apparent magnitude in the $J_{125}$ band.
Middle panel: histograms of the number counts of dropouts in our study (red) and previous studies: \citet{ouchi2009} (green), \citet{bowler2014} (yellow), and \citet{bouwens2015} (blue).
Bottom panel: best-fit Schechter function (black line) with the $1\sigma$ error (gray region) and the luminosity functions derived by this work (red circles), \citet{bouwens2014} (blue circles), \citet{bowler2014} (yellow circles), \citet{ouchi2009} (green circles), \citet{finkelstein2015} (open up-triangles), \citet{laporte2016} (open down-triangles), \citet{schenker2013} (open squares), \citet{atek2015a} (crosses), and \citet{livermore2016} (open diamonds).
The horizontal axis presents the absolute magnitude in the $J_{125}$ band.
	}
	\label{fig:fit7}
\end{figure}

\begin{figure}
	\epsscale{1.2}
	\plotone{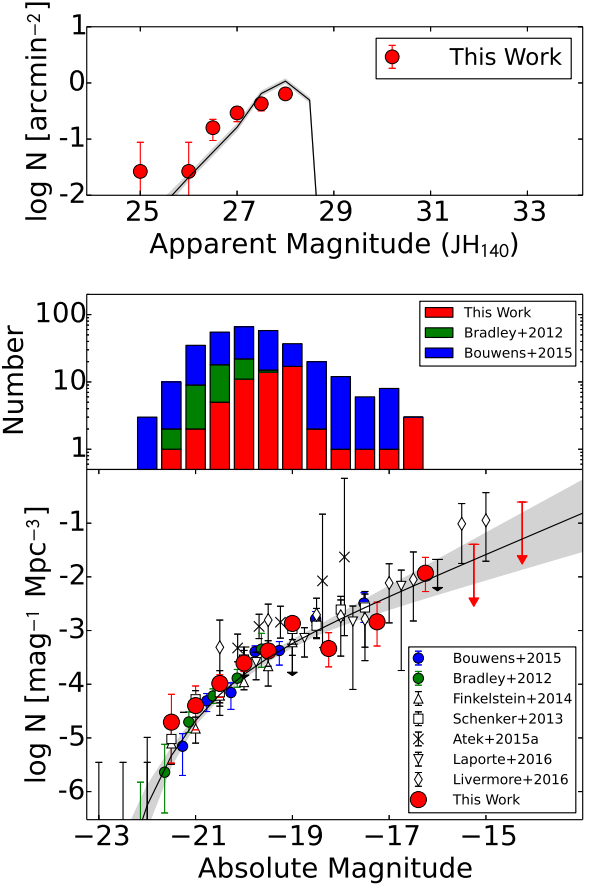}
	\caption{Same as Figure \ref{fig:fit7}, but for $z\sim8$.
The horizontal axes in the top and bottom panels present the apparent and intrinsic magnitudes in the $JH_{140}$ band, respectively.
The results of \citet{bradley2012} are shown with the green histogram (moddle panel) and circles (bottom panel).
In the bottom panel, we also plot the result of \citet{livermore2016} with the open diamonds.
	}
	\label{fig:fit8}
\end{figure}

\begin{figure}
	\epsscale{1.2}
	\plotone{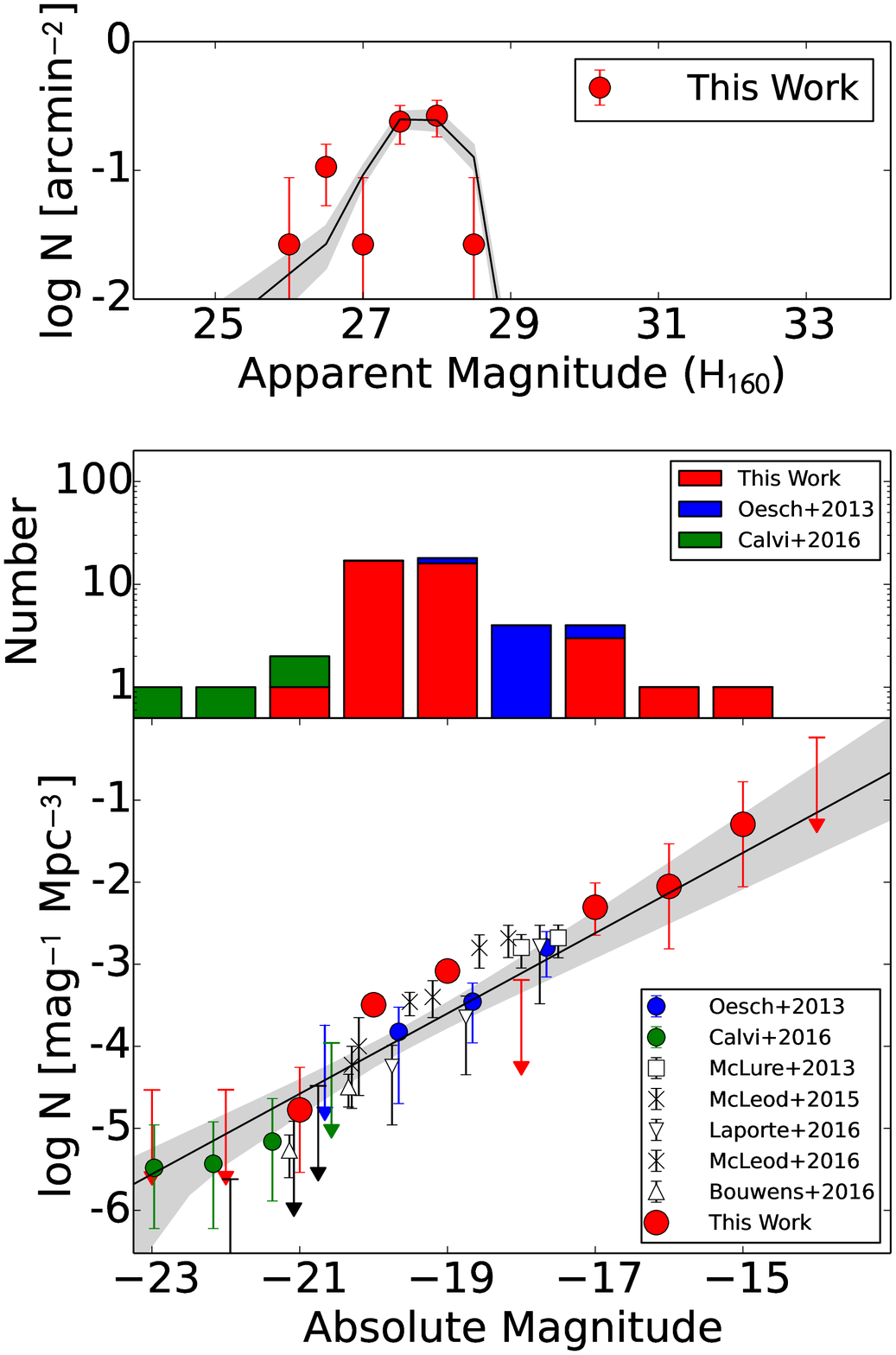}
	\caption{Same as Figure \ref{fig:fit7}, but for $z\sim9$.
The horizontal axes in the top and bottom panels present the apparent and intrinsic magnitudes in the $H_{160}$ band, respectively.
The blue (green) histogram and circles denote the results of \citet{oesch2013} \citep{calvi2016}, respectively.
In the bottom panel, we also plot the results of \citet{mclure2013} (open squares), \citet{mcleod2015,mcleod2016} (crosses), and \citet{bouwens2016d} (open triangles).
	}
	\label{fig:fit9}
\end{figure}

\begin{figure}
	\epsscale{1.2}
	\plotone{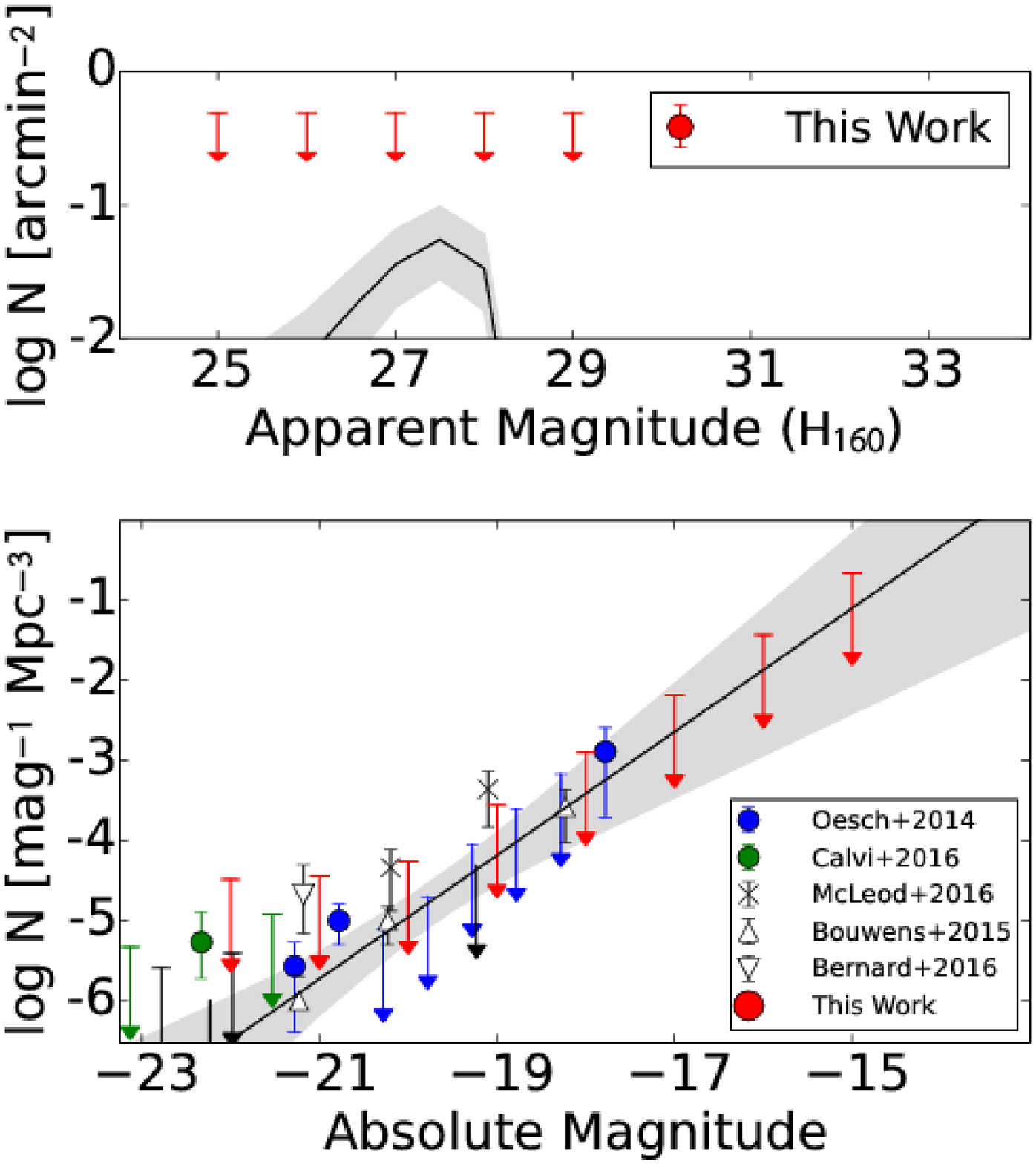}
	\caption{Same as Figure \ref{fig:fit7}, but for $z\sim10$.
The horizontal axes in the top and bottom panels present the apparent and intrinsic magnitudes in the $H_{160}$ band, respectively.
The blue (green) circles denote the results of \citet{oesch2014} \citep{calvi2016}, respectively.
In the bottom panel, we also plot the results of \citet{bernard2016} (open down-triangles), \citet{mcleod2016} (crosses), and \citet{bouwens2015} (open up-triangles).
	}
	\label{fig:fit10}
\end{figure}

The best-fit parameters are shown in Table \ref{lf_parameters}.
We find that the uncertainties in $M_*$ and $\phi_*$ are considerably large due to a degeneracy between the two parameters when all parameters are variable.
Plotted in Figure~\ref{fig:alpha} is the faint-end slope $\alpha$ as a function of redshift.
Our results indicate that the best-fit values of $\alpha$ are about $-2$ at $z\sim6-7$ to $10$, which are steeper than those at lower redshift (e.g. $\alpha\sim-1.6$ at $z\sim4$ in \citealt{bouwens2015}).
We show the fitting results at $z\sim 6-7$, $8$, $9$, and $10$ in Figures \ref{fig:fit7}, \ref{fig:fit8}, \ref{fig:fit9}, and \ref{fig:fit10}, respectively. 
The top and bottom panel present the observed number densities and the best-fit luminosity functions in the image plane and the source plane, respectively.
We also plot the results of previous blank-field surveys \citep{bouwens2015,bowler2014,ouchi2009,schenker2013,bradley2012,finkelstein2015,mclure2013,oesch2013,calvi2016} and recent HFF results in other studies \citep{laporte2016,atek2015a,mcleod2016}.
The best-fit parameters are consistent with those in previous studies.
In the top panel of Figure~\ref{fig:fit7}, there may be an excess in the observed surface number density at $J_{125} = 28.5$. The reason for this excess is not clear, although using a size--luminosity relation which gives smaller sizes at faint magnitudes may reduce this excess.
At $z\sim8$, the observed number densities at the bright end are slightly larger than the number densities in the simulation.
It is probably due to the existence of an overdense region of $z\sim8$ dropouts in the Abell 2744 cluster field.
We discussed the properties of the overdensity in \citet{ishigaki2016} (see also \citealt{atek2015b} and \citealt{zheng2014}).
At $z\sim10$, although we detect no galaxies, we can place a constraint on the luminosity function from the non-detection.
Based on the best-fit parameters where only $\phi_*$ is variable, $\sim1.4$ galaxies are expected to be detected in the HFF fields.
The middle panels of Figures \ref{fig:fit7}--\ref{fig:fit10} show histograms of the number of the dropouts.
It is seen that our samples push the magnitude limits of the luminosity functions significantly by up to $\sim3$ magnitude.

\citet{bouwens2016b} claim that the accuracy of UV luminosity functions is limited  
due to the uncertainty of mass models.
In order to estimate the uncertainty of mass models with the method we use for the luminosity function at $z\sim6-7$,
we check the magnification factors of the dropout candidates with five different mass models
from Bradac (v2 for Abell 2744, v3 for MACS0416, and v1 for MACS0717, MACS1149, AbellS1063, and Abell370; \citealt{wang2015}, \citealt{hoag2016}, see also \citealt{bradac2009}), 
CATS (v3.1 for Abell 2744, v3.1 for MACS0416, v1 for MACS0717, MACS1149, AbellS1063, and Abell370; \citealt{jauzac2015}), 
Sharon (v3 for Abell 2744 and MACS0416, v2 for MACS0717, v2.1 for MACS1149, v1 for Abell370 and AbellS1063; \citealt{johnson2014}), 
Williams (v3 for Abell 2744, v3.1 for MACS0416, v1 for MACS0717, MACS1149, AbellS1063, and Abell370), 
and Zitrin (LTM v1 for all the six clusters; \citealt{zitrin2009}, \citealt{zitrin2013}).
We obtain the UV luminosity functions at $z\sim6-7$ with these five different mass models.
Figure \ref{fig:Malpha_mu} shows the best-fit Schechter parameters and the 68\% and 95\% confidence levels.
We confirm that the difference of the best-fit parameters are $\Delta \phi_\ast \sim 0.2$, $\Delta M_\ast \sim 0.2$, and $\Delta \alpha \sim 0.1$, which are comparable to the statistical errors of these parameters (Table \ref{lf_parameters}).
Magnification factor of each dropout could be different from others in different mass model (see also Figure 11 of \citealt{ishigaki2015}),
but the values of the best-fit Schechter parameters are almost the same in all the mass models.

\begin{deluxetable*}{llll}
\tabletypesize{\scriptsize}
\tablecaption{Best-fit Schechter parameters of luminosity functions\label{lf_parameters}}
\tablewidth{0pt}
\tablehead{
		\colhead{Reference} & \colhead{$M_*$}  & \colhead{$\log \phi_*$ [Mpc$^{-3}$]} & \colhead{$\alpha$}
}
\startdata
\sidehead{$z \sim 6-7$}
	This Work & $-20.89^{+0.17}_{-0.13}$ & $-3.78^{+0.15}_{-0.15}$ & $-2.15^{+0.08}_{-0.06}$ \\
	\citet{atek2015a} & $-20.89^{+0.60}_{-0.72}$ & $-3.54^{+0.48}_{-0.45}$ & $-2.04^{+0.17}_{-0.13}$ \\
	\citet{bouwens2015} & $-20.87\pm0.26$ & $-3.53^{+0.24}_{-0.23}$ & $-2.06\pm0.13$ \\
	\citet{laporte2016} & $-20.33^{+0.37}_{-0.47}$ & $-3.43^{+0.12}_{-0.15}$ & $-1.91^{+0.26}_{-0.27}$ \\
	\citet{livermore2016} & $-20.800^{+0.058}_{-0.049}$ & $-3.667^{+0.044}_{-0.041}$ & $-2.07^{+0.04}_{-0.04}$ \\
	\citet{kawamata2017} & $-20.73^{+0.46}_{-0.81}$ &  \multicolumn{1}{c}{---} & $-1.86^{+0.17}_{-0.18}$ \\
\sidehead{$z \sim 8$}
	This Work & $-20.35^{+0.20}_{-0.30}$ & $-3.60^{+0.15}_{-0.30}$ & $-1.96^{+0.18}_{-0.15}$ \\
	\citet{bradley2012} & $-20.26^{+0.29}_{-0.34}$ & $-3.37^{+0.26}_{-0.29}$ & $-1.98^{+0.23}_{-0.22}$ \\
	\citet{bouwens2015} & $-20.63\pm0.36$ & $-3.68\pm0.32$ & $-2.02\pm0.23$ \\
	\citet{laporte2016} & $-20.32^{+0.49}_{-0.26}$ & $-3.52^{+0.58}_{-0.44}$ & $-1.95^{+0.43}_{-0.40}$ \\
	\citet{livermore2016} & $-20.742^{+0.195}_{-0.152}$ & $-3.784^{+0.145}_{-0.145}$ & $-2.02^{+0.08}_{-0.07}$ \\
	\citet{kawamata2017} & $-20.73$ (fixed) &  \multicolumn{1}{c}{---} & $-1.80^{+0.22}_{-0.30}$ \\
\sidehead{$z \sim 9$}
	This Work & $-51.39^{+18.51}_{-44.73}$ & $-19.42^{+7.73}_{-24.15}$ & $-2.22^{+0.26}_{-0.17}$ \\
	This Work & $-20.35$ (fixed) & $-3.88^{+0.08}_{-0.12}$ & $-1.96$ (fixed) \\
	\citet{oesch2013} & $-18.8\pm0.3$ & $-2.94$ (fixed) & $-1.73$ (fixed) \\
	\citet{laporte2016} & $-20.45$ (fixed) & $-4.15^{+0.15}_{-0.24}$ & $-2.17^{+0.41}_{-0.43}$ \\
	\citet{mcleod2016} & $-20.1$ (fixed) & $-3.62^{+0.08}_{-0.10}$ & $-2.02$ (fixed) \\
	\citet{kawamata2017} & $-20.73$ (fixed) &  \multicolumn{1}{c}{---} & $-1.59^{+0.19}_{-0.18}$ \\
\sidehead{$z \sim 10$}
	This Work & $-59.71^{+29.90}_{-9.61}$ & $-35.60^{+23.73}_{-5.93}$ & $-2.93^{+0.83}_{-0.41}$ \\
	This Work & $-20.35$ (fixed) & $-4.60^{+0.14}_{-0.22}$ & $-1.96$ (fixed) \\
	\citet{oesch2014} & $-20.12$ (fixed) & $-4.27\pm0.21$ & $-2.02$ (fixed \\
	\citet{bouwens2015} & $-20.92$ (fixed) & $-5.10^{+0.18}_{-0.20}$ & $-2.27$ (fixed) \\
	\citet{mcleod2016} & $-20.1$ (fixed) & $-3.90^{+0.13}_{-0.20}$ & $-2.02$ (fixed) 
\enddata
\end{deluxetable*}

\begin{figure}
	\epsscale{1.2}
	\plotone{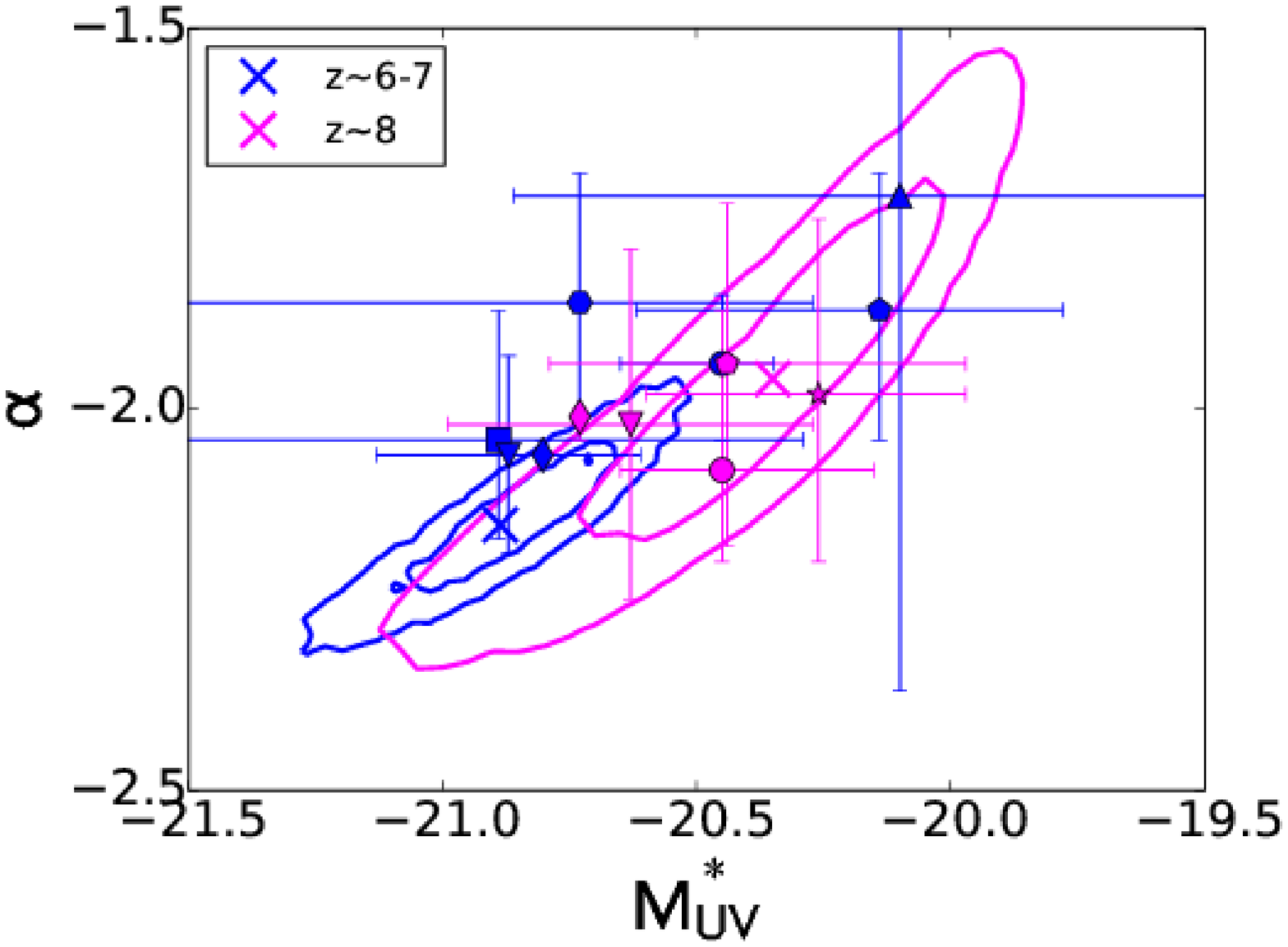}
	\caption{Contours of Schechter parameters $M_\ast$ and $\alpha$ indicating the 68\% and 95\% confidence levels.
In this plot, blue (magenta) symbols an lines denote $z\sim6-7$ ($z\sim8$).
The values of the best-fit Schechter parameters are shown with the cross symbols.
We plot the best-fit parameters from previous studies: \citet{ishigaki2015} (circles), \citet{atek2015a} (square), \citet{bouwens2015} (down-triangles), \citet{ouchi2009} (up-triangle), \citet{schenker2013} (pentagons), \citet{livermore2016} (diamonds), \citet{bradley2012} (star), and \citet{kawamata2017} (hexagon).
	}
	\label{fig:Malpha}
\end{figure}

\begin{figure}
	\epsscale{1.2}
	\plotone{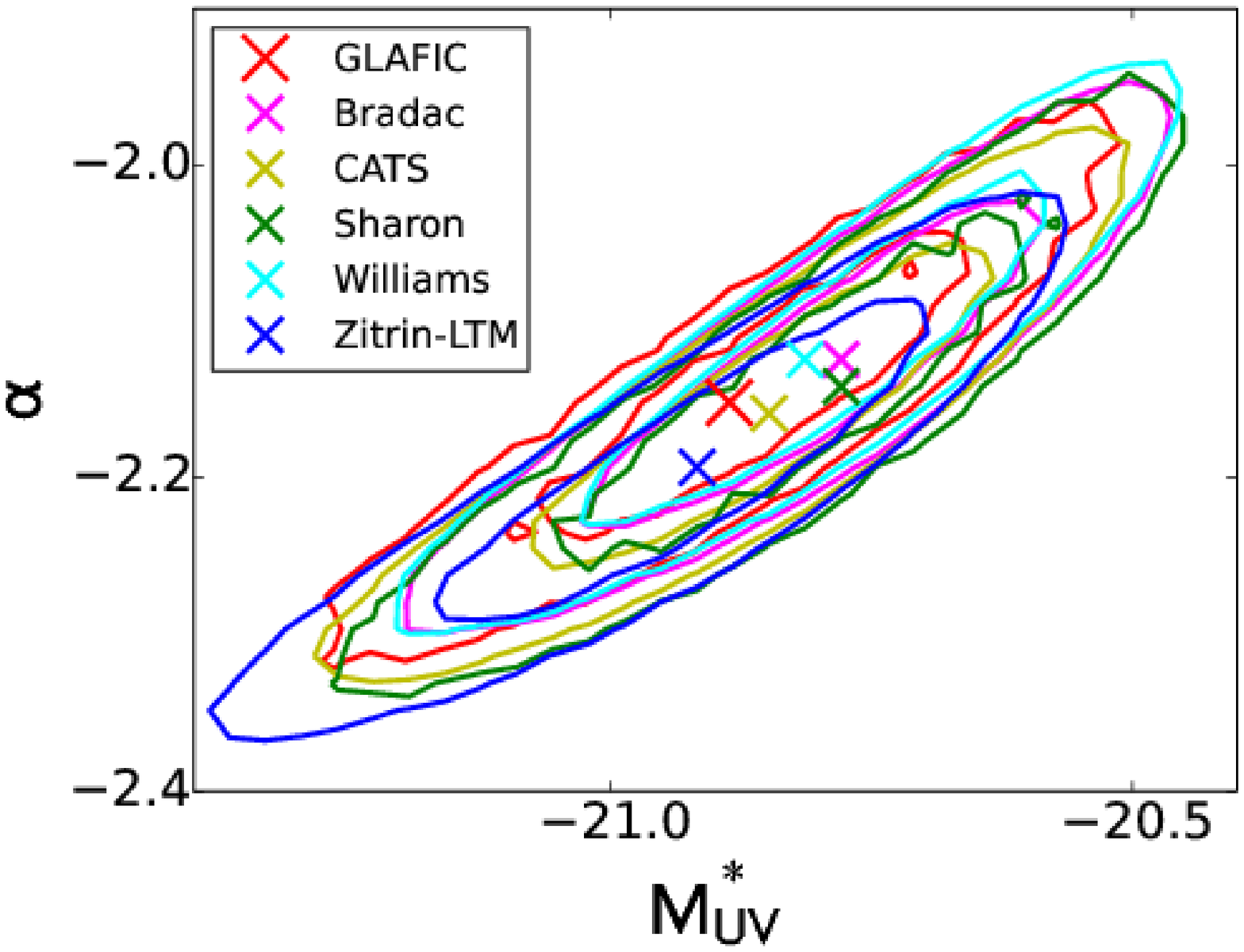}
	\caption{Contours of Schechter parameters $M_\ast$ and $\alpha$ at $z\sim6-7$ with different mass models; GLAFIC (red), Bradac(magenta), CATS (yellow), Sharon (green), Williams (cyan), and Zitrin-LTM (blue).
The values of the best-fit Schechter parameters are shown with the cross symbols.
	}
	\label{fig:Malpha_mu}
\end{figure}

\section{Discussion}\label{sec:Discussion}

In the previous section, we derive the UV luminosity functions at $z\sim6-10$ from the HFF data and the previous blank field data.
In this section, we estimate the UV luminosity densities $\rho_\mathrm{UV}$ from the luminosity functions which constrain the physical properties of ionizing sources and reionization.

\subsection{Evolution of the UV Luminosity Density}\label{sec:rho}

\begin{figure}
	\epsscale{1.2}
	\plotone{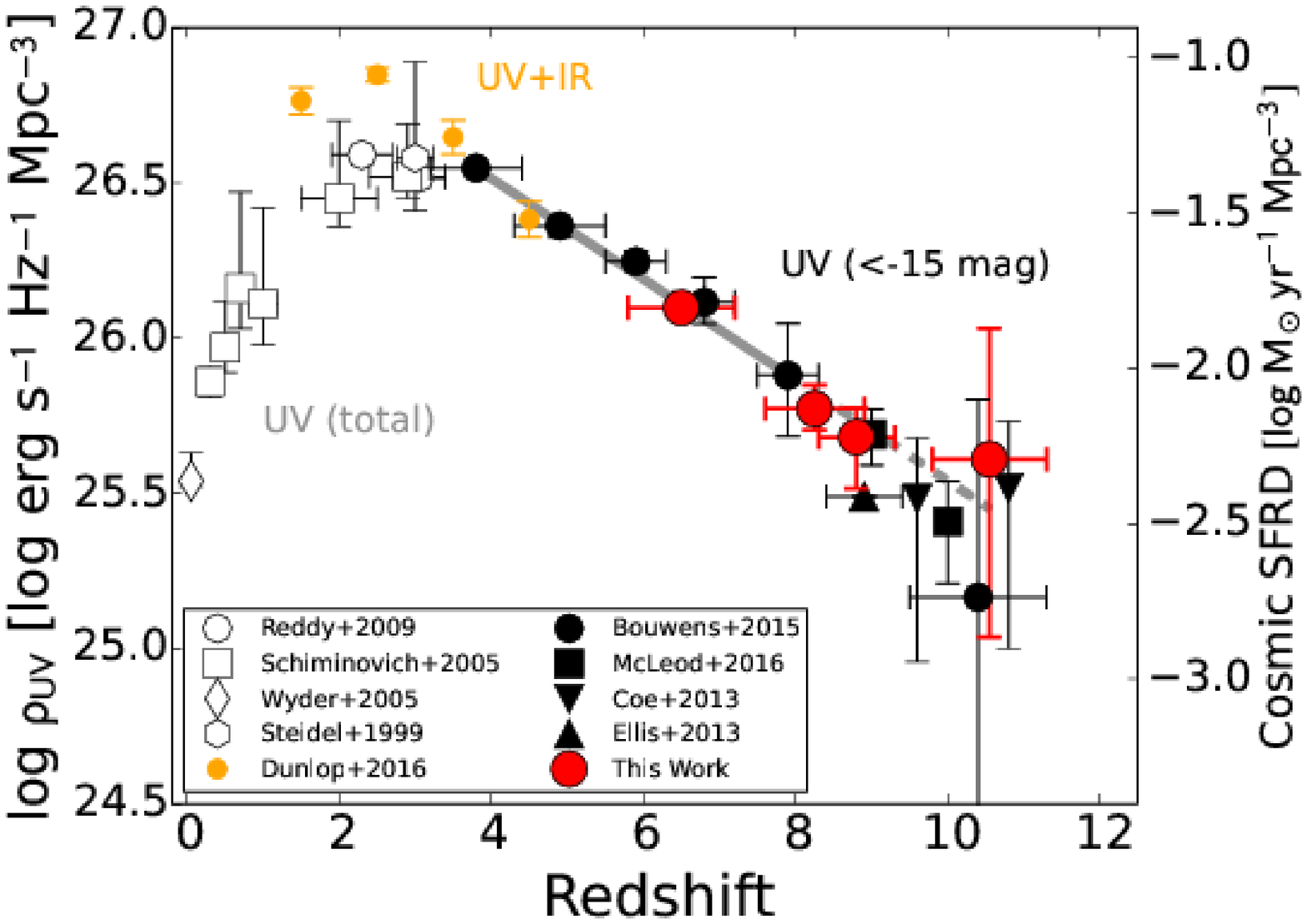}
	\plotone{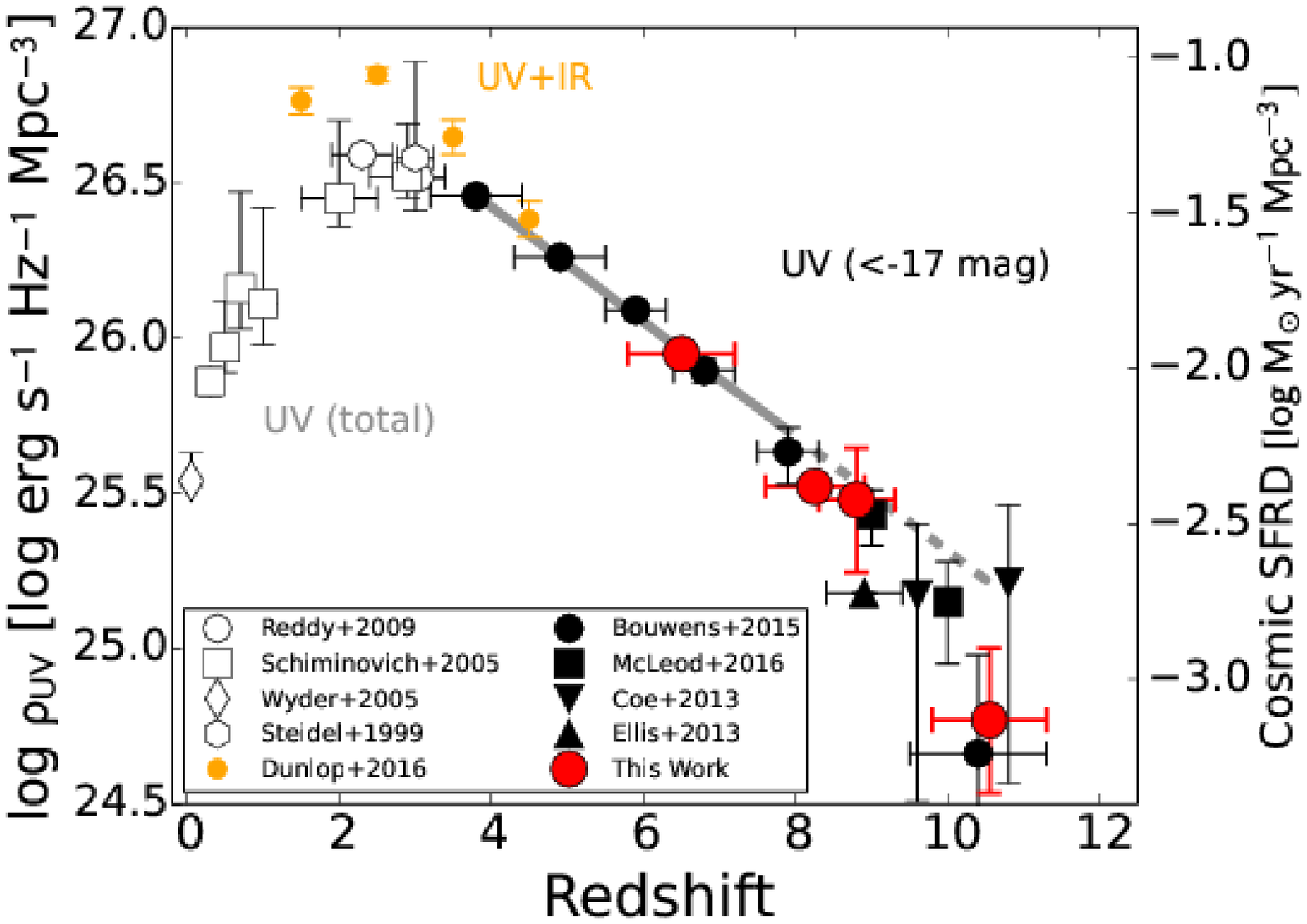}
	\caption{UV luminosity densities $\rho_\mathrm{UV}$ calculated with $M_{\mathrm{trunc}}=-15$ (\textit{top}) and $M_{\mathrm{trunc}}=-17$ (\textit{bottom}).
We show $\rho_\mathrm{UV}$ taken from this work (red circles), \citet{bouwens2015} (black filled circles), \citet{mcleod2016} (black filled squares), \citet{coe2013} (black filled down-triangles), and \citet{ellis2013} (black filled up-triangle).
We also plot total $\rho_\mathrm{UV}$ from \citet{steidel1999} (open hexagon), \citet{wyder2005} (open diamond), \citet{schiminovich2005} (open squares), and \citet{reddy2009} (open circles).
The orange circles denote the cosmic SFRDs derived with $\rho_\mathrm{UV}$ and IR luminosity densities \citep{dunlop2016}.
The gray solid and dashed lines show a linear fitting to samples in this work and \citet{bouwens2015} at $z\sim4-8$ and its extrapolation to $z\sim8-10$, respectively.
	}
	\label{fig:rho}
\end{figure}

The value of $\rho_\mathrm{UV}$ is calculated with the following equation:
\begin{eqnarray}
\rho_\mathrm{UV}(z) = \int^{M_\mathrm{trunc}}_{-\infty} \Phi(M)L(M)dM,
\label{eq:rho}
\end{eqnarray}
where $L(M)$ is the UV luminosity corresponding to the UV magnitude $M$. 
The value of $M_\mathrm{trunc}$ is the truncation magnitude of the UV luminosity function,
i.e., there are no galaxies with the magnitudes fainter than $M_\mathrm{trunc}$.
We calculate $\rho_\mathrm{UV}$ at $z\sim6-7$, $8$, $9$, and $10$ using the best-fit Schechter functions obtained in Section \ref{sec:fitting}.
Figure \ref{fig:rho} shows $\rho_\mathrm{UV}$ calculated from Equation (\ref{eq:rho}) with $M_\mathrm{trunc} = -15.0$ and $M_\mathrm{trunc} = -17.0$.
The right axis in Figure \ref{fig:rho} presents cosmic star formation rate densities (SFRDs) from a given luminosity density that are estimated from $\rho_\mathrm{UV}$ based on Equation (2) of \citet{madau1998}.

For comparisons, we also show the data plots of $\rho_\mathrm{UV}$ with $M_\mathrm{trunc} = -15.0$ and $M_\mathrm{trunc} = -17.0$ at $z\sim4-12$ that are taken from previous studies \citep{bouwens2015,mcleod2016,coe2013,ellis2013}.
In Figure \ref{fig:rho}, we include total $\rho_\mathrm{UV}$ at $z\sim0-3$ \citep{steidel1999,wyder2005,schiminovich2005,reddy2009},
and cosmic SFRDs at $z\sim1-4$ based on $\rho_\mathrm{UV}$ ($M_\mathrm{trunc} = -15.0$) and IR luminosity densities (the sensitivity corresponding to $\simeq 30$ $\mu$Jy) \citep{dunlop2016}.
Our results are consistent with these previous studies at $z\sim6-10$.

Our results support a non-accelerated decline of $\rho_\mathrm{UV}$ toward high redshifts in the case of $M_\mathrm{trunc} = -15.0$ (see also \citealt{mcleod2016}).
We check whether we can approximate the evolution of $\log \rho_\mathrm{UV}$ from $z=4$ to $10$ as a linear function.
We fit the $\log \rho_\mathrm{UV}$ from this work and \citet{bouwens2015} with linear and quadratic functions.
The Akaike Information Criterion corrected for small data sets \citep[AIC\textit{c}; ][]{sugiura1978} for the linear function is smaller by $\Delta \mathrm{AIC}c = 3.69$ than that for the quadratic function, suggesting a non-accelerated $\rho_\mathrm{UV}$ evolution.

However, we find an accelerated decline when a brighter truncation magnitude of $M_\mathrm{trunc} = -17.0$ is adopted, which is closer to $M_\mathrm{trunc} = -17.7$ adopted in \citet{oesch2013}, who support an accelerated decline.
In Figure~\ref{fig:rho}, we present fitting results to $\rho_\mathrm{UV}$ in this work and \citet{bouwens2015} with a linear function at $z\sim4-8$ and its extrapolation toward $z\sim10$.
This indicates the trends of non-accelerated and accelerated declines in the cases of $M_\mathrm{trunc} = -15.0$ and $-17.0$, respectively.
One possible explanation for this dependence on $M_\mathrm{trunc}$ is the evolution of the faint-end slope toward steeper value at higher redshifts (see Figure~\ref{fig:alpha}).
The faint-end slope steepening makes the $\rho_\mathrm{UV}$ difference (between the cases of $M_\mathrm{trunc} = -15.0$ and $-17.0$) larger at $z\sim10$ than at $z\sim8$. 
The redshift evolution of $M_\ast$ toward fainter magnitudes (see Figure~\ref{fig:Malpha}) would also contribute to this larger increases of $\rho_\mathrm{UV}$ at higher redshifts.

On the other hand, our previous study in \citet{ishigaki2015} has claimed that $\rho_\mathrm{UV}$ shows an accelerated decrease beyond $z\sim8$.
This can be because they adopt a brighter truncation magnitude of $M_\mathrm{trunc} = -17.0$.
Another possible explanation for the difference is cosmic variance.
\citet{ishigaki2015} only use the data of Abell 2744 cluster and parallel fields.
In these fields, the galaxy density at $z\sim8$ is $\sim0.1$ dex higher than the average of total fields \citep{ishigaki2016},
and the one at $z\sim9$ is $\sim0.1$ dex lower than the average.

\subsection{Preperties of the Ionizing Sources}\label{sec:ionizing_sources}

In this section, we calculate the ionized hydrogen fraction $Q_\mathrm{HII}$ and the Thomson scattering optical depth $\tau_\mathrm{e}$,
basically in the same manner as that described in Section 5 of \citet{ishigaki2015}.

The value of $Q_\mathrm{HII}$ is calculated with the following equation (e.g. \citealt{robertson2013}):
\begin{eqnarray}
		\dot{Q}_\mathrm{H_{II}} = \frac{\dot{n}_\mathrm{ion}}{\average{n_\mathrm{H}}} - \frac{Q_\mathrm{H_{II}}}{t_\mathrm{rec}}.
\label{eq:Q}
\end{eqnarray}
In the first term of Equation (\ref{eq:Q}), $\dot{n}_\mathrm{ion}$ and $\average{n_\mathrm{H}}$ denote the production rate of ionizing photons and the mean hydrogen number density, respectively.
$\dot{n}_\mathrm{ion}$ and $\average{n_\mathrm{H}}$ are defined by the following equations:
\begin{eqnarray}
		\dot{n}_{{\rm ion}} &=& \int^{M_{\rm trunc}}_{-\infty} f_{\rm esc}(M_{\rm UV}) \xi_{\rm ion}(M_{\rm UV}) \Phi(M_{\rm UV}) L(M_{\rm UV}) dM_{\rm UV}\nonumber \\
		&\equiv & \left< f_{\rm esc} \xi_{\rm ion} \right> \rho_{\rm UV}, \label{eq:nion}\\
		\average{n_{\rm H}} &=& \frac{X_{\rm p} \Omega_{\rm b} \rho_{\rm c}}{m_{\rm H}}, \label{eq:nH}
\end{eqnarray}
where $X_\mathrm{p}$ is the primordial mass fraction of hydrogen, $\rho_\mathrm{c}$ is the critical density, and $m_\mathrm{H}$ is the mass of the hydrogen atom.
$f_\mathrm{esc}$ is the fraction of the number of escaping ionizing photons to those produced in a galaxy. 
$\xi_\mathrm{ion}$ is the numerical factor that converts a UV luminosity density to the ionizing photon emission rate of a star-forming galaxy.
Note that $f_\mathrm{esc}$ and $\xi_\mathrm{ion}$ appear in the product form in Equation (\ref{eq:nion}).
Here we assume that $f_\mathrm{esc}$ and $\xi_\mathrm{ion}$ do not depend on $M_\mathrm{UV}$.
We present the magnitude-averaged value of the product of $f_\mathrm{esc}$ and $\xi_\mathrm{ion}$ as $\left< f_{\rm esc} \xi_{\rm ion} \right>$.
In the second term of Equation (\ref{eq:Q}), $t_\mathrm{rec}$ represents the averaged gas recombination time:
\begin{eqnarray}
		t_{{\rm rec}} = \frac{1}{C_{\rm H_{II}} \alpha_{\rm B}(T) (1 + Y_{\rm p} / 4X_{\rm p}) \average{n_{\rm H}} (1+z)^3}, \label{eq:trec}
\end{eqnarray}
where $\alpha_\mathrm{B}$ is the case-B hydrogen recombination coefficient, $T$ is the IGM temperature, and $Y_\mathrm{p}$ is the primordial helium mass fraction.
$C_\mathrm{H_{II}} \equiv \left<n_\mathrm{H_{II}}^2\right> / \left<n_\mathrm{H_{II}}\right>^2$ is a clumping factor, where $n_\mathrm{H_{II}}$ is the local number density of ionized hydrogen.
In this work, we use the following equation:
\begin{eqnarray}
C_\mathrm{H_{II}} = 2.9 \times \left[\frac{(1+z)}{6}\right]^{-1.1},
\end{eqnarray}
which was obtained from the hydrodynamical $+$ $N$-body simualtions in \citet{shull2012}.
The value of the clumping factor is not accurately known observationally, although many theoretical studies suggest similar values of clumping factor, $C_\mathrm{H_{II}} \sim 1-6$, in the epoch of reionization \citep{sokasian2003,Iliev2006,finlator2012}.

The value of $\tau_\mathrm{e}$ is calculated with the following equation \citep{kuhlen2012}:
\begin{eqnarray}
		\begin{aligned}
		\tau_e(z) = \int^z_0 \frac{c(1+z')^2}{H(z')}Q_{{\rm H_{II}}} \sigma_{\rm T} \average{n_{\rm H}} \\
		\times (1+\eta Y_{\rm p}/4X_{\rm p}) dz'  \label{eq:tau_e},
		\end{aligned}
\end{eqnarray}
where $H(z)$ is the Hubble parameter, $\sigma_\mathrm{T}$ is the Thomson scattering cross section, and $c$ is the speed of light. 
We assume that helium is singly ionized (corresponding to $\eta = 1$) at $z > 4$ and doubly ionized (corresponding to $\eta = 2$) at $z \leq 4$, following \citet{kuhlen2012}.
Recently, \citet{planck2016} have obtained $\tau_\mathrm{e} = 0.058\pm0.012$ from the measurement of $TT$ power spectrum and $EE$ polarization of CMB.
The value of $\tau_\mathrm{e}$ from \citet{planck2016} is smaller than the one obtained in previous observations by nine-year WMAP ($\tau_\mathrm{e} = 0.088 \pm 0.013$, \citealt{bennett2013,hinshaw2013}) and Planck 2014 ($\tau_\mathrm{e} = 0.092 \pm 0.013$, \citealt{planck2014}).

\begin{figure*}
	\epsscale{1.2}
	\plotone{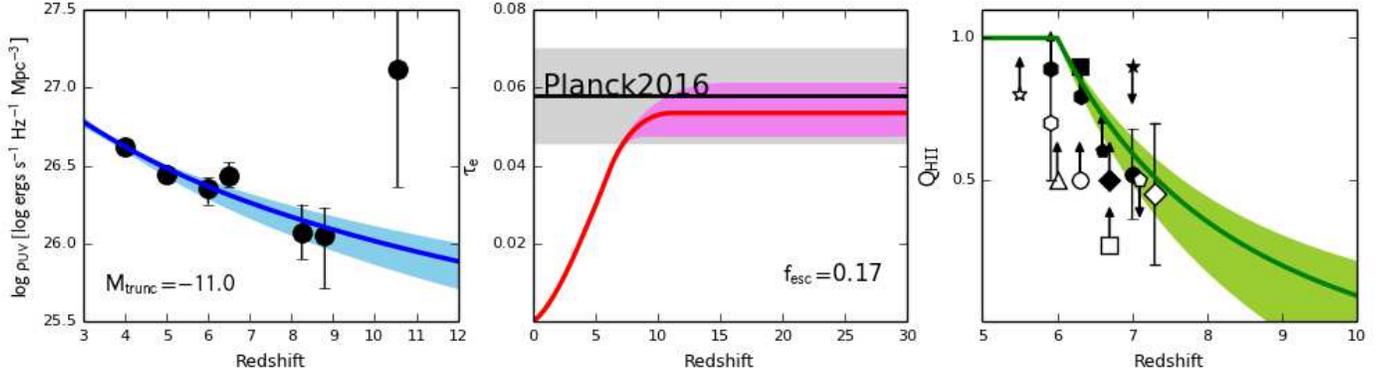}
	\caption{
Left panel: $\rho_\mathrm{UV}$ calculated with $M_\mathrm{trunc} = -11.0$. 
The black circles present $\rho_\mathrm{UV}$ from the best-fit luminosity functions at $z\sim7-10$ (this work) and those at $z\sim4-6$ \citep{bouwens2015}.
The blue line and the light blue shade denote the best-fit function of $\rho_\mathrm{UV}$ and the $1\sigma$ error, respectively, calculated with Equation (\ref{eq:rho_z}) (see text for the best-fit parameters).
The up-scattering $\rho_\mathrm{UV}(z)$ at $z\sim10$ due to the large uncertainty in $\alpha$ does not affect the fitting result, because the uncertainty in $\rho_\mathrm{UV}(z)$ is also large.
Middle panel: $\tau_\mathrm{e}$ integrating from $z=0$ to a redshift $z$.
The red line and the magenta shade represent $\tau_\mathrm{e}(z)$ and the $1\sigma$ error, respectively, that are consistent with $\rho_\mathrm{UV}(z)$ shown with the blue line in the left panel.
The black line and the gray region show the values of $\tau_\mathrm{e}$ and its $1\sigma$ error, respectively, obtained by \citet{planck2016}.
Right panel: evolution of $Q_\mathrm{HII}$.
The green line and the light green shade present $Q_\mathrm{HII}$ and the $1\sigma$ error, respectively, that agree with $\rho_\mathrm{UV}(z)$ shown with the blue line in the left panel based on Equation \ref{eq:Q}.
The symbols denote constraints of $Q_\mathrm{HII}$ from \citet{ota2008} (filled circle), \citet{konno2014} (open diamond), \citet{carilli2010} (filled square), \citet{bolton2011} (filled star), \citet{mcquinn2008} (open circle), \citet{ouchi2010} (filled diamond), \citet{mcquinn2007} (filled pentagon), \citet{mesinger2010} (open triangle), \citet{mcgreer2011} (open star), \citet{mcquinn2007,mesinger2008,dijkstra2011} (open pentagon), \citet{chornock2013,chornock2014} (filled hexagons), \citet{totani2014} (open hexagon), and \citet{patel2010} (open square).
	}
	\label{fig:plot_tile}
\end{figure*}

We assume the following two functional form of $\rho_\mathrm{UV}$.
One is the four-parameter function from \citet{madau2014} (see also \citealt{robertson2015}):
\begin{eqnarray}
\rho_\mathrm{UV}(z) = a_p\frac{(1+z)^b_p}{1+\left[(1+z)/c_p\right]^d_p},
\label{eq:rho_z}
\end{eqnarray}
which has free parameters of $a_p$, $b_p$, $c_p$, and $d_p$.
The other is a logarithmic double power law function used in \citet{ishigaki2015}:
\begin{eqnarray}
\rho_\mathrm{UV}(z) = \frac{2 \rho_\mathrm{UV}^{\ast}}{10^{a(z-z_\ast)}+10^{b(z-z_\ast)}},
\label{eq:rho2_z}
\end{eqnarray}
which also has four free parameters; $\rho_\mathrm{UV}^{\ast}$, $z_\ast$, $a$, and $b$.
We perform $\chi^2$ fitting to the observational data of $\rho_\mathrm{UV}$, $Q_\mathrm{HII}$, and $\tau_\mathrm{e}$.
In each parameter space, we adopt one of the two functional forms of $\rho_\mathrm{UV}(z)$ whose minimum $\chi^2$ is smaller than the one of the other.
We use the $\rho_\mathrm{UV}$ data points at $z\sim6-7$, $8$, $9$, and $10$ (this work) and $z\sim4$, $5$, and $6$ \citep{bouwens2015} that are presented in the left panel of Figure \ref{fig:plot_tile}.
Although $\rho_\mathrm{UV}$ at $z\sim10$ scatters upward due to the large uncertainty in $\alpha$, this does not significantly affect the fitting result because of its large uncertainty.
We also use the $Q_\mathrm{HII}$ data plotted in the right panel of Figure \ref{fig:plot_tile} from \citet{bolton2011,carilli2010,chornock2013,chornock2014,dijkstra2011,konno2014,mcgreer2011,mcquinn2007,mcquinn2008,mesinger2008,mesinger2010,ouchi2010,ota2008,patel2010,totani2006,totani2014}.
In addition to these observational constraints, we compare $\tau_\mathrm{e} = 0.058\pm0.012$ \citep{planck2016} with the value of $\tau_\mathrm{e}$ at $z=30$ calculted from Equation (\ref{eq:tau_e}).
There are six free parameters in the $\chi^2$ fit, $\left< f_{\rm esc} \xi_{\rm ion} \right>$, $M_\mathrm{trunc}$, and the four parameters in the function of $\rho_\mathrm{UV}(z)$.
The parameter range of $\left< f_{\rm esc} \xi_{\rm ion} \right>$ is $0$ to $10^{25.34}$ [erg$^{-1}$ Hz],
where we assume that $0 < f_\mathrm{esc} < 1$ and $\log \xi_\mathrm{ion}/[\mathrm{erg}^{-1}\ \mathrm{Hz}] = 25.34$  \citep{bouwens2016c}.
The parameter range of $M_\mathrm{trunc}$ is $-16$ to $-10$;
$M_\mathrm{trunc} = -16$ mag corresponds to the detection limit of current observations,
and $M_\mathrm{trunc} = -10$ mag is the magnitude of minimium halos which have star forming galaxies predicted by \citet{faucher2011}.

We calculate the total $\chi^2$ value by summing up the $\chi^2$ of the data points of $\rho_\mathrm{UV}$, $Q_\mathrm{HII}$, and $\tau_\mathrm{e}$, and derive the best-fit parameters.
The best-fit parameters are $\log a_p=25.9$, $b_p=3.7$, $c_p=2.3$, $d_p=5.5$, $\left< \log f_{\rm esc} \xi_{\rm ion} \right>=24.57$ and $M_\mathrm{trunc}=-11.0$,
and the $\chi^2$ is $9.28$ for $12$ degrees of freedom.
Figure \ref{fig:plot_tile} shows the best-fit functions of $\rho_\mathrm{UV}$, $\tau_\mathrm{e}$, and $Q_\mathrm{HII}$.
These best-fit functions agree well with the data points of the observations.
This result is in contrast with the conclusion of our previous study of \citet{ishigaki2015}, which claim that no parameter set can reproduce both the $\rho_\mathrm{UV}$ evolution and the value of $\tau_\mathrm{e}$.
The main reason for this difference is that \citet{ishigaki2015} use the value $\tau_\mathrm{e} = 0.091^{+0.013}_{-0.014}$ from \citet{planck2014}, which is significantly larger than the latest result used in our study ($\tau_\mathrm{e} = 0.058\pm0.012$; \citealt{planck2016}).
Another reason is that this study supports a non-accelerated decline of $\rho_\mathrm{UV}$ toward high redshift as explained in Section \ref{sec:rho}.
In this study, the small value of $\tau_\mathrm{e}$ and a non-accelerated decline of $\rho_\mathrm{UV}$ alleviate the tension between $\tau_\mathrm{e}$ and $\rho_\mathrm{UV}$ claimed by \citet{ishigaki2015}.

\begin{figure}
	\epsscale{1.2}
	\plotone{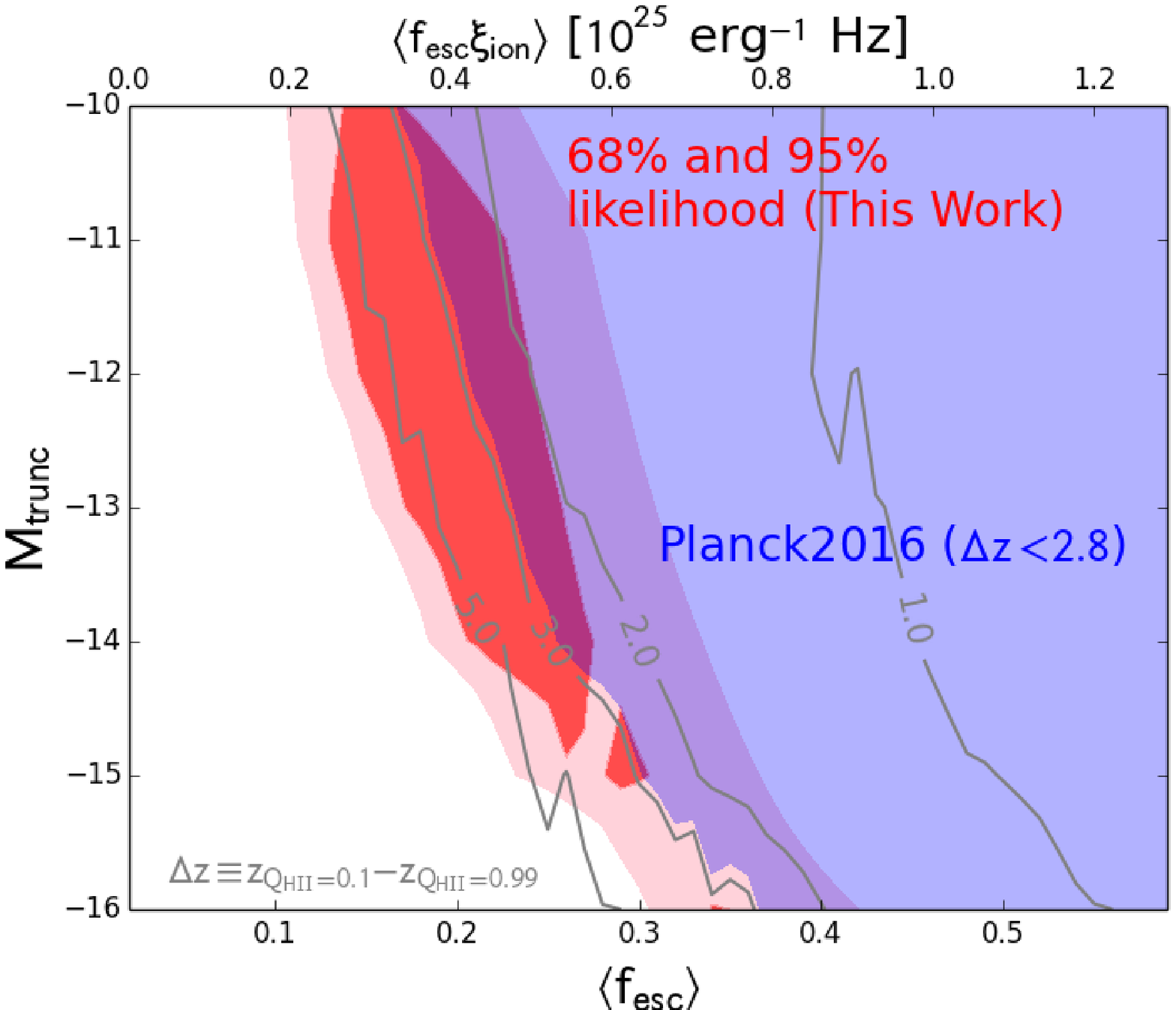}
	\caption{
Contours at 68\% (pink shade) and 95\% (magenta shade) confidence levels of $\left< f_{\rm esc} \xi_{\rm ion} \right>$ and $M_\mathrm{trunc}$ parameters.
The upper horizontal axis represents $\left< f_{\rm esc} \xi_{\rm ion} \right>$, and the lower horizontal axis denotes the average escape fraction $\left< f_{\rm esc} \right>$ under the assumption of $\log \xi_\mathrm{ion}/[\mathrm{erg}^{-1}\ \mathrm{Hz}] = 25.34$ \citep{bouwens2016c}.
The blue shade shows the 68\% confidence contour calculated with the constraints of $\Delta z$ from \citet{planck2016}.
The solid lines represent $\Delta z = 1.0$, $2.0$, $3.0$, and $5.0$.
	}
	\label{fig:fesc_Mlim}
\end{figure}

Figure \ref{fig:fesc_Mlim} presents the $68\%$ and $95\%$ confidence intervals on the $\left< f_{\rm esc} \xi_{\rm ion} \right>$--$M_\mathrm{trunc}$ plane with megenta contours.
The best-fit value and the 68\% interval are $\left< \log f_{\rm esc} \xi_{\rm ion} \right> = 24.57^{+0.15}_{-0.08}$ and $M_\mathrm{trunc} > -14.0$.
If we assume $\log \xi_\mathrm{ion}/[\mathrm{erg}^{-1}\ \mathrm{Hz}] = 25.34$ obtained in \citet{bouwens2016c}, we place upper and lower limits on the escape fraction, $f_\mathrm{esc} < 0.24$ and $f_\mathrm{esc} > 0.14$, respectively.
These constraints of $f_\mathrm{esc}$ and $M_\mathrm{trunc}$ are mainly driven by the observational constraints of $Q_\mathrm{HII}$.
We also derive the length of the reionization period $\Delta z$
that is defined by the period bracketed by two redshifts whose $Q_\mathrm{HII}$ values are $0.1$ and $0.99$.
Figure \ref{fig:fesc_Mlim} shows the contours of $\Delta z$ predicted by our $\chi^2$ fitting results.
The contours suggest $\Delta z = 3.9 ^{+2.0}_{-1.6}$ in the $68\%$ confidence interval on the $\left< f_{\rm esc} \xi_{\rm ion} \right>$--$M_\mathrm{trunc}$ plane.
\citet{planck2016} obtain $\Delta z < 2.8$ from kSZ constraints and a reionization model, which is indicative of the relatively sharp reionization history.
The blue shade in Figure \ref{fig:fesc_Mlim} presents the $68\%$ confidence interval constrained from the result of \citet{planck2016} on the $\left< f_{\rm esc} \xi_{\rm ion} \right>$--$M_\mathrm{trunc}$ plane.
Our constraints on $\Delta z$ are complimentary to those of \citet{planck2016}, giving both the upper and lower limit of $\Delta z$.
The combination of the constraints from our and \citet{planck2016} studies suggest $\Delta z \sim 2-3$, which suggests the moderately sharp reionization history.

\section{Summary}\label{sec:Summary}

In this study, we have produced catalogs of dropout galaxies at $z\sim6-10$ using the all cluster and parallel fields data taken by the HFF program.
Using our new mass models, we have conducted Monte-Carlo simulations, and estimated the number densities of dropout galaxies at $z\sim6-10$. 
We then derive the UV luminosity densities and discuss the properties of ionizing sources and reionization.
The results in this study are summarized below:
\begin{enumerate}
\item With the dropout selection technique, we identify $\sim 400$ star-forming galaxies over the redshift range $z\sim6-10$, which include 350 $i$-dropout, 66 $Y$-dropout, and 40 $YJ$-dropout candidates. 
We do not find $J$-dropout candiates in the HFF data.
The number of dropout candidates in our catalogs is six times larger than the one obtained in our previous study \citep{ishigaki2015} that use one set of the cluster and parallel field data.
\item The faint end slope of UV luminosity functions has a very steep value $\alpha\sim-2$.
The UV luminosity densities calculated from the UV luminosity functions are consistent with those in previous studies. 
Our results support the evolutionary trend of a non-accelerated decline in the UV luminosity densities beyond $z\sim8$ in the case of $M_\mathrm{trunc}=-15$, consistent with the conclusions of \citet{mcleod2016}, while an accelerated decline in the case of a brighter truncation magnitude of $M_\mathrm{trunc}=-17$, consistent with the conclusions of \citet{oesch2013, bouwens2015, ishigaki2015, oesch2015}.
\item Using the standard analytical reionization model, we calculate the physical parameters related to reionization, such as the Thomson scattering optical depth $\tau_\mathrm{e}$ and the ionized hydrogen fraction $Q_\mathrm{HII}$.
The values of $\tau_\mathrm{e}$ and $Q_\mathrm{HII}$ are consistent with the latest result of \citet{planck2016} ($\tau_\mathrm{e} = 0.0058\pm0.0012$) and previous studies of $Q_\mathrm{HII}$ obtained with Ly$\alpha$ absorptions found in QSOs, GRBs, and galaxies.
We obtain the constraints on the parameters of ionizing sources. 
The best-fit value and the $68\%$ confidence intervals of the averaged value of the escape fraction and truncation magnitude are $f_{\rm esc} = 0.17^{+0.07}_{-0.03}$ and $M_\mathrm{trunc} > -14.0$, respectively, under the assumption of $\log \xi_\mathrm{ion}/[\mathrm{erg}^{-1}\ \mathrm{Hz}] = 25.34$. 
The results suggest that the length of the reionization period is $\Delta z = 3.9 ^{+2.0}_{-1.6}$
which is consistent with the $\Delta z$ estimate from the kinetic Sunyaev-Zel’dovich effect in \citet{planck2016}.
\end{enumerate}

\acknowledgments
We are grateful to Akio Inoue, Masayuki Umemura, Masanori Iye, Kentaro Nagamine, Takatoshi Shibuya for useful comments and discussions.
This research was supported by the Munich Institute for Astro- and Particle Physics (MIAPP) of the DFG cluster of excellence "Origin and Structure of the Universe".
This work is supported by World Premier International Research Center Initiative (WPI Initiative), MEXT, Japan, and KAKENHI (JP15H02064, JP26800093, and JP15H05892) Grant-in-Aid for Scientific Research (A) through Japan Society for the Promotion of Science (JSPS).
M.I. acknowledges support by Grant-in-Aid for JSPS Research Fellow (JP16J03727)
and an Advanced Leading Graduate Course for Photon Science grant.
R.K. acknowledges support by Grant-in-Aid for JSPS Research Fellow (JP16J01302).

{\it Facilities:} \facility{{\sl HST} (WFC3, ACS)}


\bibliographystyle{apj}
\bibliography{ishigaki2017}

\begin{appendix}
\section{Lists of Dropout Candidates at $z\sim6-7$, $8$, and $9$}

\setlength{\tabcolsep}{6pt}
\LongTables
\begin{deluxetable*}{lccccccccc}
\tabletypesize{\scriptsize}
\tablecaption{Dropout candidates at $z\sim 6-7$\label{candidates7}}
\tablewidth{0pt}
\tablehead{
		\colhead{ID} & \colhead{R.A. (J2000)} & \colhead{Dec (J2000)} & \colhead{$i_{814}-Y_{105}$} & \colhead{$Y_{105}-J_{125}$} & \colhead{$J_{125}$\tablenotemark{a}} & \colhead{$\mu$\tablenotemark{b}} & \colhead{Photo-$z$} & \colhead{Reference\tablenotemark{c}}
}
\startdata
\sidehead{Abell2744 Cluster}
HFF1C-$2251$-$4556$& $3.593804$& $-30.415447$& $   2.68 \pm    0.36$& $   0.05 \pm    0.05$& $   26.10 \pm    0.04$ & $  3.64$ & $  6.62_{-0.75}^{+0.75}$ & (1), (2), (7)\\
HFF1C-$1695$-$4527$& $3.570654$& $-30.414659$& $   1.33 \pm    0.10$& $   0.11 \pm    0.05$& $   26.22 \pm    0.04$ & $  1.47$ & $  6.00_{-0.69}^{+0.69}$ & (1), (2)\\
HFF1C-$2549$-$3119$& $3.606222$& $-30.386644$& $   1.07 \pm    0.08$& $   0.08 \pm    0.05$& $   26.26 \pm    0.04$ & $  1.55$ & $  5.80_{-0.67}^{+0.67}$ & (1), (2), (7)\\
HFF1C-$1930$-$4181$& $3.580452$& $-30.405043$ & $      >    2.36$& $   0.17 \pm    0.09$& $   26.61  \pm    0.07$ & $  4.47$ & $  6.79_{-0.76}^{+0.76}$ & (1), (2), (7) \\
HFF1C-$2516$-$4570$& $3.604865$& $-30.415839$& $   0.92 \pm    0.14$& $   0.06 \pm    0.08$& $   26.67 \pm    0.06$ & $  2.84$ & $  5.64_{-0.65}^{+0.65}$ & \nodata\\
HFF1C-$2348$-$3454$& $3.597834$& $-30.395961$ & $      >    2.13$& $   0.27 \pm    0.11$& $   26.79  \pm    0.08$ & $  3.31$ & $  7.00_{-0.78}^{+0.78}$ & (2), (7) \\
HFF1C-$2178$-$2458$& $3.590761$& $-30.379408$& $   0.96 \pm    0.14$& $  -0.16 \pm    0.10$& $   27.07 \pm    0.08$ & $  1.68$ & $  5.84_{-0.67}^{+0.67}$ & (7)\\
HFF1C-$2047$-$3526$& $3.585321$& $-30.397958$ & $      >    1.83$& $   0.13 \pm    0.15$& $   27.19  \pm    0.11$ & $  3.23$ & $  6.75_{-0.76}^{+0.76}$ & (2), (7), (10) \\
\enddata
\tablenotetext{a}{Total magnitudes $m_\mathrm{tot}$ estimated with the aperture correction $m_\mathrm{tot} = m_\mathrm{AP} - 0.82$.}
\tablenotetext{b}{Magnifications calculated with mass models in \citet{kawamata2017}.}
\tablenotetext{c}{References. (1)\citet{atek2014}, (2)\citet{atek2015a}, (3)\citet{coe2015}, (4)\citet{laporte2014}, (5)\citet{laporte2015}, (6)\citet{laporte2016}, (7)\citet{livermore2016}, (8)\citet{mcleod2016}, (9)\citet{oesch2015} (10)\citet{zheng2014}, (11)\citet{zitrin2014}}
\end{deluxetable*}

\setlength{\tabcolsep}{6pt}
\LongTables
\begin{deluxetable*}{lccccccccc}
\tabletypesize{\scriptsize}
\tablecaption{Dropout candidates at $z\sim 8$\label{candidates8}}
\tablewidth{0pt}
\tablehead{
		\colhead{ID} & \colhead{R.A. (J2000)} & \colhead{Dec (J2000)} & \colhead{$Y_{105}-J_{125}$} & \colhead{$J_{125}-JH_{140}$} & \colhead{$JH_{140}$\tablenotemark{a}} & \colhead{$\mu$\tablenotemark{b}} & \colhead{Photo-$z$} & \colhead{Reference\tablenotemark{c}}
}
\startdata
\sidehead{Abell2744 Cluster}
HFF1C-$2508$-$2496$& $3.604518$& $-30.380467$& $   1.17 \pm    0.07$& $   0.04 \pm    0.04$& $   25.92 \pm    0.03$ & $  1.31$ & $  8.00_{-0.88}^{+0.88}$ & (1), (2), (3), (4), (10) \\
HFF1C-$2481$-$2561$& $3.603378$& $-30.382254$& $   1.27 \pm    0.15$& $   0.14 \pm    0.08$& $   26.64 \pm    0.05$ & $  1.40$ & $  8.20_{-0.90}^{+0.90}$ & (2), (3), (7), (10) \\
HFF1C-$2306$-$3089$& $3.596091$& $-30.385833$& $   1.30 \pm    0.16$& $   0.06 \pm    0.08$& $   26.68 \pm    0.06$ & $  2.04$ & $  8.18_{-0.90}^{+0.90}$ & (2), (3), (7), (10) \\
HFF1C-$2555$-$2515$& $3.606461$& $-30.380996$& $   1.12 \pm    0.14$& $  -0.18 \pm    0.09$& $   26.95 \pm    0.07$ & $  1.33$ & $  7.96_{-0.88}^{+0.88}$ & (2), (3), (7), (10) \\
HFF1C-$2492$-$2561$& $3.603859$& $-30.382263$& $   1.92 \pm    0.37$& $   0.16 \pm    0.11$& $   26.99 \pm    0.07$ & $  1.40$ & $  8.40_{-0.92}^{+0.92}$ & (2), (3), (7), (8), (10) \\
HFF1C-$2557$-$2513$& $3.606577$& $-30.380924$& $   1.10 \pm    0.24$& $   0.22 \pm    0.13$& $   27.17 \pm    0.08$ & $  1.32$ & $  7.88_{-6.13}^{+0.87}$ & (2), (10) \\
HFF1C-$2135$-$2432$& $3.588980$& $-30.378668$& $   1.13 \pm    0.19$& $  -0.19 \pm    0.13$& $   27.30 \pm    0.10$ & $  1.69$ & $  7.94_{-0.88}^{+0.88}$ & (2), (3), (10) \\
HFF1C-$2495$-$2562$& $3.603997$& $-30.382304$& $   1.05 \pm    0.32$& $   0.13 \pm    0.19$& $   27.61 \pm    0.12$ & $  1.40$ & $  7.90_{-6.08}^{+0.87}$ & (10) \\
\enddata
\tablenotetext{a}{Total magnitudes $m_\mathrm{tot}$ estimated with the aperture correction $m_\mathrm{tot} = m_\mathrm{AP} - 0.82$.}
\tablenotetext{b}{Magnifications calculated with mass models in \citet{kawamata2017}.}
\tablenotetext{c}{References. (1)\citet{atek2014}, (2)\citet{atek2015a}, (3)\citet{coe2015}, (4)\citet{laporte2014}, (5)\citet{laporte2015}, (6)\citet{laporte2016}, (7)\citet{livermore2016}, (8)\citet{mcleod2016}, (9)\citet{oesch2015} (10)\citet{zheng2014}, (11)\citet{zitrin2014}}
\end{deluxetable*}

\LongTables
\begin{deluxetable*}{lccccccccc}
\tabletypesize{\scriptsize}
\tablecaption{Dropout candidates at $z\sim 9$\label{candidates9}}
\tablewidth{0pt}
\tablehead{
		\colhead{ID} & \colhead{R.A. (J2000)} & \colhead{Dec (J2000)} & \colhead{$YJ-JH_{140}$} & \colhead{$JH_{140}-H_{160}$} & \colhead{$H_{160}$\tablenotemark{a}} & \colhead{$\mu$\tablenotemark{b}} & \colhead{Photo-$z$} & \colhead{Reference\tablenotemark{c}}
}
\startdata
\sidehead{Abell2744 Cluster}
HFF1C-$2481$-$2561$& $3.603380$& $-30.382255$& $   0.78 \pm    0.08$& $  -0.05 \pm    0.07$& $   26.69 \pm    0.05 $ & $  1.40$ & $  8.19_{-0.90}^{+0.90}$ & (2), (3), (7), (10) \\
HFF1C-$2492$-$2561$& $3.603859$& $-30.382262$& $   1.12 \pm    0.11$& $   0.19 \pm    0.09$& $   26.79 \pm    0.06 $ & $  1.40$ & $  8.40_{-0.92}^{+0.92}$ & (1), (2), (3), (7), (10) \\
HFF1C-$2220$-$4053$& $3.592512$& $-30.401486$ & $      >    1.21$& $   0.55 \pm    0.30$& $   27.38  \pm    0.17$ & $ 14.13$ & $  9.54_{-7.31}^{+1.03}$ & (9), (11) \\
\sidehead{Abell2744 Parallel}
HFF1P-$5363$-$4016$& $3.473469$& $-30.400460$& $   0.96 \pm    0.18$& $   0.08 \pm    0.15$& $   27.44 \pm    0.10 $ & \nodata & $  8.30_{-6.63}^{+0.91}$ & \nodata \\
HFF1P-$5364$-$3024$& $3.473522$& $-30.384024$ & $      >    1.34$& $   0.26 \pm    0.22$& $   27.77  \pm    0.13$ & \nodata & $  8.73_{-6.34}^{+0.95}$ & (8) \\
HFF1P-$5733$-$3462$& $3.488893$& $-30.396183$& $   1.65 \pm    0.26$& $  -0.23 \pm    0.18$& $   27.77 \pm    0.14 $ & \nodata & $  8.72_{-0.95}^{+0.95}$ & (8) \\
HFF1P-$5386$-$2074$& $3.474445$& $-30.368728$ & $      >    1.80$& $  -0.26 \pm    0.27$& $   28.17  \pm    0.21$ & \nodata & $  8.89_{-6.87}^{+0.97}$ & \nodata \\
\sidehead{MACS0416 Cluster}
HFF2C-$1151$-$4540$& $64.047984$& $-24.081671$& $   0.97 \pm    0.06$& $   0.00 \pm    0.06$& $   26.50 \pm    0.04 $ & $  1.44$ & $  8.39_{-0.92}^{+0.92}$ & (3), (5), (8) \\
\enddata
\tablenotetext{a}{Total magnitudes $m_\mathrm{tot}$ estimated with the aperture correction $m_\mathrm{tot} = m_\mathrm{AP} - 0.82$.}
\tablenotetext{b}{Magnifications calculated with mass models in \citet{kawamata2017}.}
\tablenotetext{c}{References. (1)\citet{atek2014}, (2)\citet{atek2015a}, (3)\citet{coe2015}, (4)\citet{laporte2014}, (5)\citet{laporte2015}, (6)\citet{laporte2016}, (7)\citet{livermore2016}, (8)\citet{mcleod2016}, (9)\citet{oesch2015} (10)\citet{zheng2014}, (11)\citet{zitrin2014}}
\end{deluxetable*}

\end{appendix}

\end{document}